\documentclass[acmtog]{acmart}
\setcopyright{none}
\renewcommand\footnotetextcopyrightpermission[1]{}
\usepackage{booktabs} 
\usepackage[ruled]{algorithm2e} 

\usepackage{wrapfig}
\usepackage{overpic}
\usepackage{multirow}

\SetAlFnt{\small}
\SetAlCapFnt{\small}
\SetAlCapNameFnt{\small}
\SetAlCapHSkip{0pt}

\begin{document}
\title{Filmsticking++: Rapid Film Sticking for Explicit Surface Reconstruction}

\author{Pengfei Wang}
    \affiliation{%
    \institution{Shandong University}
    \city{Jinan}
    \country{China}}

\author{Jian Liu}
    \affiliation{%
    \institution{Shenyang University of Technology}
    \city{Shenyang}
    \country{China}}

\author{Qiujie Dong}
    \affiliation{%
    \institution{The University of Hong Kong}
    \city{Hong Kong}
    \country{China}}

\author{Shiqing Xin}       
    \affiliation{%
    \institution{Shandong University}
    \city{Qingdao}
    \country{China}}
    
\author{Yuanfeng Zhou}
    \affiliation{%
    \institution{Shandong University}
    \city{Jinan}
    \country{China}}

\author{Changhe Tu}
    \affiliation{%
    \institution{Shandong University}
    \city{Qingdao}
    \country{China}}

\author{Caiming Zhang}
    \affiliation{%
    \institution{Shandong University}
    \city{Jinan}
    \country{China}}

\author{Wenping Wang}
    \affiliation{%
    \institution{Texas A\&M University}
    \state{Texas}
    \country{United States of America}}

\begin{abstract}

Explicit surface reconstruction aims to generate a surface mesh that exactly interpolates a given point cloud. This requirement is crucial when the point cloud must lie non-negotiably on the final surface to preserve sharp features and fine geometric details. However, the task becomes substantially challenging with low-quality point clouds, due to inherent reconstruction ambiguities compounded by combinatorial complexity.
A previous method using filmsticking technique by iteratively compute restricted Voronoi diagram to address these issues, ensures to produce a watertight manifold, setting a new benchmark as the state-of-the-art (SOTA) technique.
Unfortunately, RVD-based filmsticking is inability to interpolate all points in the case of deep internal cavities, it is necessary to employ numerous cycles of filmsticking and sculpting to progressively approach the real geometry, resulting in very likely is the generation of faulty topology.
The cause of this issue is that RVD-based filmsticking has inherent limitations due to Euclidean distance metrics.
In deep internal cavities, the dominated regions of sites far from the guiding surface may be obstructed by those of other sites.
In this paper, we extend the filmsticking technique, named Filmsticking++.
Filmsticking++ reconstructing an explicit surface from points without normals.
On one hand, Filmsticking++ break through the inherent limitations of Euclidean distance by employing a weighted-distance-based Restricted Power Diagram, which guarantees that all points are interpolated.
On the other hand, we observe that as the guiding surface increasingly approximates the target shape, the external medial axis is gradually expelled outside the guiding surface. Building on this observation, we propose placing virtual sites inside the guiding surface to accelerate the expulsion of the external medial axis from its interior.
Additionally, in the thin-plate model, a site typically dominate multiple disconnected regions. 
To maintain manifoldness, only one of these regions can be retained. 
However, filmsticking employs a heuristic that selects the region closest to the site, ignoring topological considerations, which easily confuses the two sides of the thin-plate model. To address this, we propose a topology-aware manifold fix strategy.
To summarize, contrary to the SOTA method, Filmsticking++ demonstrates multiple benefits, including decreases computational cost, improved robustness and scalability.

\end{abstract}



\begin{teaserfigure}
  \centering
  \includegraphics[width=\textwidth]{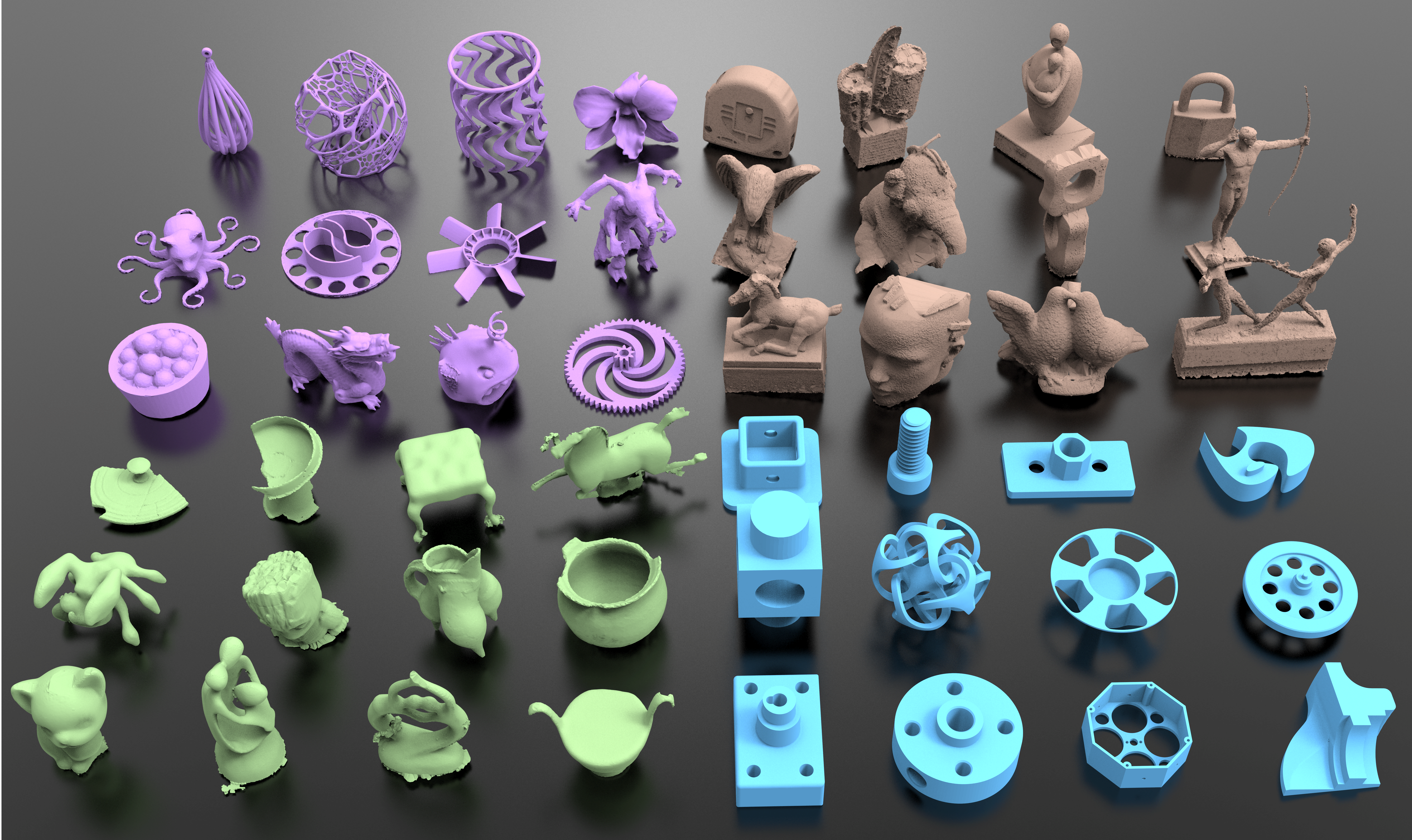}\\
  \caption{A gallery of surfaces reconstruction using our method Filmsticking++, with four kinds of point clouds input: nonuniform sampling on thingi10k~\cite{2019thingi10k} (purple), real scan on public data set~\cite{huang2022surface} and ~\cite{EPFL} (brown), real scan on ~\cite{wang2022restricted} and SHINING 3D Einscan SE scanner (green), and accurately coincide with the surface sampling on ABC public data set~\cite{koch2019abc} (blue). In this paper, we introduce a rapid filmsticking~\cite{wang2022restricted} technique for explicit surface reconstruction, termed Filmsticking++. Compared with filmsticking, our method stands out for its theoretical elegance and simplicity in algorithmic design, resulting in a more robust implementation. 
  }
  \label{fig:teaser}
\end{teaserfigure}

\maketitle
\section{Introduction}
Surface reconstruction~\cite{Berger2016A, lim2014surface, berger2013benchmark} involves recovering the underlying geometry, typically represented as a polygonal mesh~\cite{botsch2007geometric, botsch2010polygon}, from partial information such as point clouds. 
It is useful in a variety of fields, including computer graphics~\cite{aliaga20123d}, computer vision~\cite{okura20223d}, and 3D modeling~\cite{caumon2009surface}. 
Surface reconstruction remains a hot research topic in recent years.

There are two main categories of reconstruction methods: implicit reconstruction approaches~\cite{10.1145/383259.383266, kazhdan2013screened} and explicit ones~\cite{bernardini1999ball, 10.1007/s00371-003-0217-z, digne2011scale}. Implicit approaches involve inferring an implicit function and then extracting the zero-level iso-surface as the reconstructed result, while explicit approaches aim to directly connect the points into a triangle mesh. Both types of approaches have their advantages. Implicit approaches excel at producing smooth surfaces and naturally eliminate noise to some extent, whereas explicit approaches strive to preserve as much fidelity as possible from the raw data. The primary focus of this paper is explicit surface reconstruction.

\begin{figure}[ht]
  \begin{center}
  \includegraphics[width=0.95\columnwidth]{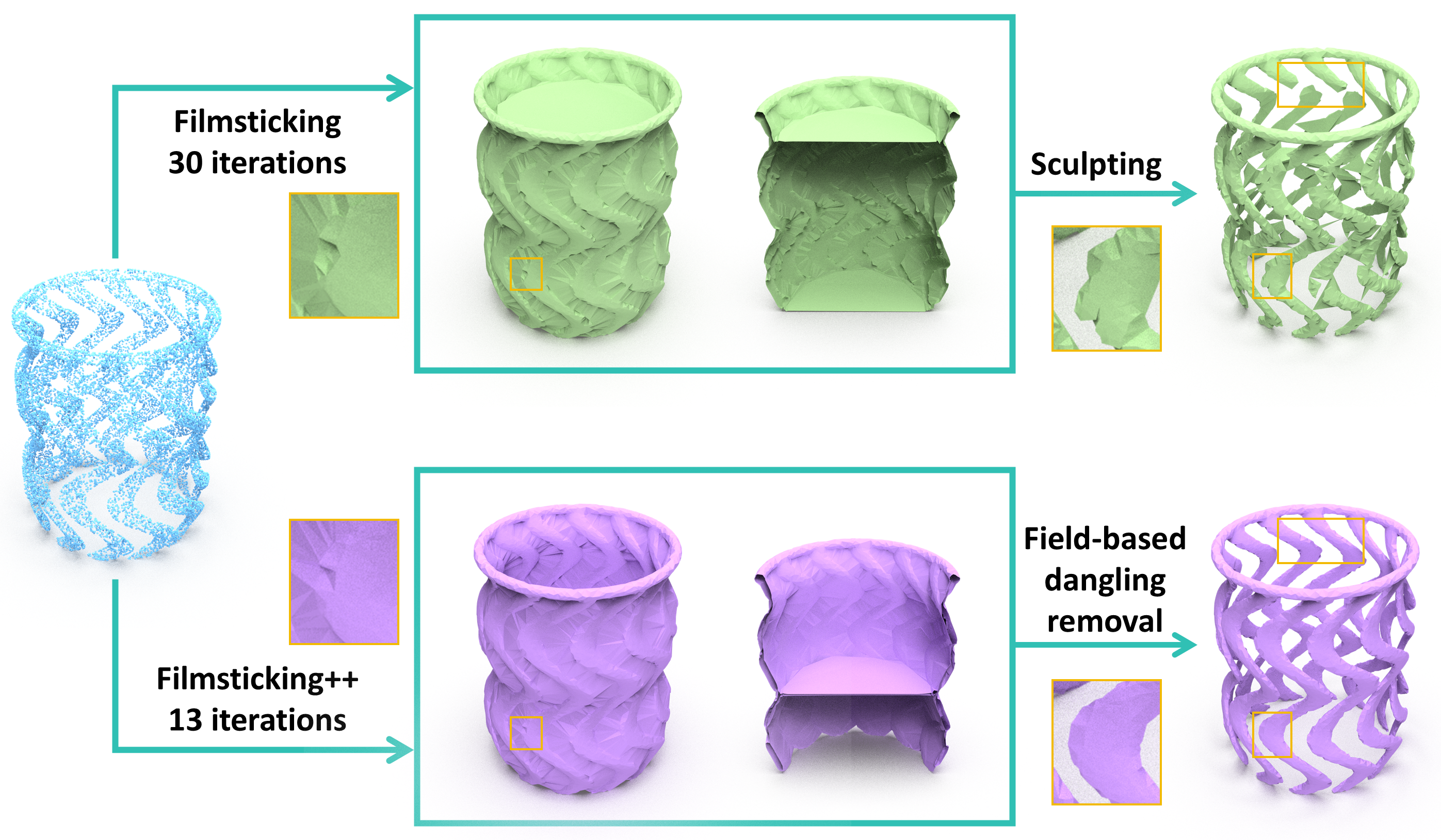}
  \vspace{-3mm}
  \end{center}
   \caption{ 
   The state-of-the-art (SOTA) method ~\cite{wang2022restricted} falls short in attracting the guiding surface to adhere to the internal cavity, resulting in the need for multiple cycles of filmsticking and sculpting operations.
   In contrast, our only use 13 Filmsticking++ and employs just one field-based dangling removal operation, significantly reduces the number of iterations and generate a high-fidelity mesh surfaces.
  }
   \label{fig:Problem_old_filmsticking}
   \vspace{-2mm}
\end{figure}

The most challenging aspect of explicit approaches lies in overcoming inherent ambiguity and combinatorial complexity. 
Traditional methods such as Ball Pivoting (BP)~\cite{bernardini1999ball} and Greedy\cite{10.1007/s00371-003-0217-z}, based on the principle of local optimality, can rapidly interpolate points one by one into a triangular mesh. However, they fail to account for global consistency and are unable to generate a topologically correct watertight manifold mesh when faced with non-uniform point cloud distributions, resulting in numerous non-manifold artifacts and holes. Learn-based interpolation methods~\cite{lei2023circnet, rakotosaona2021learning, sharp2020pointtrinet} employ soft penalties to enhance their ability to produce manifold and watertight surfaces but may struggle with data imperfections and manifold mesh generation.

Recently, ~\cite{wang2022restricted} proposed an interpolation-based approach utilizing restricted Voronoi diagrams (RVD), which guarantees the generation of a watertight manifold and has set a new state-of-the-art benchmark. Their method begins with a watertight manifold sphere as the initial guiding surface, computes the RVD on guiding surface, and updates guiding surface via restricted Delaunay triangulation (RDT). 
This core procedure, termed filmsticking, is applied iteratively, enabling the guiding surface to progressively approximate the target geometry. 
Following filmsticking, a genus‑0 mesh model is obtained. Subsequent volumetric sculpting is then performed within the regions enclosed by the guiding surface where it significantly deviates from the input data.

However, the filmsticking operation becomes inapplicable for point clouds representing shapes with deep cavities or thin plates, as it faces three key limitations: 1) the guiding surface cannot be attracted into internal cavities, 2) extensive iterative filmsticking and sculpting is needed inside cavities, and 3) the two sides of a thin plate cannot be distinguished.
When confronting the first issue, their heuristic sculpting method is prone to confusing the actual interior and exterior of the shape, leading to erroneous excision of internal geometry (see Fig.~\ref{fig:Problem_old_filmsticking}).

The research in this paper builds upon ~\cite{wang2022restricted} by introducing a more flexible filmsticking method, named Filmsticking++. The primary focus of Filmsticking++ is to enhance robustness in handling complex shapes and attract the guiding surface with fewer iterations. 

First, Voronoi diagrams offer local neighborhood properties but lack global consistency. When integrated with the guiding surface, the filmsticking operation can progressively refine these local neighborhoods while preserving global coherence. However, RVD-based filmsticking is inherently limited by its reliance on Euclidean distance metrics. In complex shapes, the dominated regions (Voronoi cells) of sites far from the guiding surface can be obstructed by those of neighboring sites, preventing them from being attracted to the surface and thus underutilizing local neighborhood information. To overcome this limitation, we propose replacing the restricted Voronoi diagram with the restricted power diagram (RPD). Our RPD-based Filmsticking++ ensures that all points are attracted to the guiding surface, fully leveraging local neighborhood properties to reconstruct the underlying geometry more accurately.

Second, we observed that filmsticking within internal cavities leads to very slow evolution of the guiding surface, as only one site may be attracted per iteration. To accelerate this process, we scatter a set of auxiliary points to serve as virtual sites inside the guiding surface, which are allowed to be attracted onto it. Once a virtual site reaches the surface, it is removed in the next iteration and cannot be attracted again. As shown in Fig.~\ref{fig:Problem_old_filmsticking}, our method significantly reduces the number of required iterations.

Finally, during the early stages of guiding surface evolution, the two sides of a thin plate can be easily confused, as illustrated in Fig.~\ref{fig:Problem_old_filmsticking}. Once such confusion occurs, filmsticking~\cite{wang2022restricted} may propagate this error through subsequent iterations, causing it to persist in the final result. To address this, we introduce a smoothness-aware manifold Voronoi diagram, which incorporates local smoothness as a geometric prior. When a site governs multiple disconnected regions, instead of using Euclidean distance, we apply a local topological smoothness criterion to select the most plausible region. This approach effectively corrects erroneous topological associations and ensures robust reconstruction of thin structures.

In addition, the guiding surface produced by our Filmsticking++ incorporates almost all input points and closely approximates the true geometry. Therefore, a cumbersome sculpting operation may not be necessary, as most field-based methods, such as Poisson reconstruction~\cite{10.5555/1281957.1281965}, can assist in obtaining the correct topology, as discussed in Section~\ref{Sculpting}.

The implementation involves addressing at least two key challenges.
First, improperly assigned weights in the RPD may cause one site’s dominated region to overwhelm those of neighboring sites, thereby disrupting the underlying Euclidean neighborhood structure. To mitigate this, we have identified and formally proven the existence of a safe weight interval that ensures points are effectively attracted to the guiding surface while preserving the original neighborhood relationships. Further details are provided in Section~\ref{RPD_GuidingSurface}.
Second, poorly positioned virtual sites can impede convergence speed. We observe that as the guiding surface approaches the target geometry, the external medial axis of the true shape is gradually expelled outward. Accordingly, we place virtual sites along the medial axis direction, aligning the strongest deformation direction of the guiding surface with the external medial axis to accelerate convergence. Additional implementation aspects are discussed in Section~\ref{VirtualSites}.

In summary, Filmsticking++ offers multiple advantages, including fewer iterations, improved robustness, and better scalability.

\section{Related Work}

Surface reconstruction from 3D point clouds is a pivotal endeavor in the field of computer graphics, with a multitude of algorithms~\cite{SurveyStructedOrUnstructed} emerging over the past four decades. 
In this section, we provide a concise overview of the prevalent algorithms, and introduce approximation methods and interpolation methods in point cloud reconstruction sequentially. 
The output of approximation methods approximates the input point cloud, while interpolation methods strictly adhere to the input point cloud, directly connecting the points to form triangular meshes.
Our method is an interpolation approach based on restricted Delaunay triangulation, and the most closely related work will be introduced at the end.

\subsection{Approximation Methods}
Approximation-based methods offer the primary advantage of generating smooth manifold triangle surfaces while also exhibiting strong robustness to noise. 
However, they also have significant limitations. First, they do not interpolate the input points, which makes it difficult to fully preserve original geometric details such as sharp features, thin-plate structures, or tubular shapes. To improve geometric fidelity, specialized techniques—such as adaptive subdivision~\cite{Fuhrmann2014Floating,2016Global,2017Visibility} or persistence diagrams~\cite{TogologyAwareRec}—must be carefully incorporated. Second, most approximation methods depend on reliable normal vectors for stable performance.
Existing approximation approaches can generally be classified into two categories: traditional methods and deep learning-based methods.

\subsubsection{Traditional approximation methods}
A substantial body of literature ~\cite{Benchmark} focuses on deducing implicit equations and subsequently extracting zero-level set surfaces. For instance, radial basis functions (RBFs) ~\cite{10.1145/383259.383266,2010Smoothing} serve as effective tools for inferring the implicit representation of surfaces. Signed-distance fields (SDFs) also play a key role in approximating the underlying geometric field. Methods such as Poisson reconstruction ~\cite{10.5555/1281957.1281965} and its screened variant ~\cite{kazhdan2013screened,Fuhrmann2014Floating} formulate an inside-outside indicator field—analogous to an SDF—as the solution to a Poisson equation. When dealing with very sparse point clouds, ~\cite{PointToMesh} proposed learning and predicting surfaces using regression forests. Additionally, novel techniques have been developed to address specific reconstruction challenges, such as those encountered in urban buildings or indoor scenes ~\cite{Yi2017Urban, 2017Modeling,2018Sparse3D}. For example, ~\cite{2018Sparse3D} introduced a 3D global matching algorithm designed to handle issues like insufficient temporal sampling or rapid camera motion.
In addition, ~\cite{2017Field} introduced a field-aligned online surface reconstruction method that eliminates the need for signed-distance computations typically required in conventional approaches, enabling an efficient, output-driven interactive scanning and reconstruction workflow. Readers interested in a broader survey of related techniques may refer to ~\cite{Berger2016A}. Other notable approximation-based strategies exist as well. For example, one line of work approximates the underlying surface using a collection of geometric primitives ~\cite{SparseImplicit2005, PolyFitRec}.
Among them, the Variational Implicit Surface Reconstruction method (VIPSS) ~\cite{huang2019variational} demonstrates strong robustness to sparse and non-uniform point distributions. However, this robustness comes at the cost of a global smoothness regularizer that can oversmooth sharp geometric features. Moreover, VIPSS has a computational complexity of $O(n^3)$, restricting its practical use to point clouds of roughly 5,000 points. To overcome this limitation, ~\cite{ju2025variational} proposed NNVIPSS, a locally reformulated version of VIPSS, which improves both runtime performance and scalability while maintaining reconstruction quality.

\subsubsection{Learning-Based approximation methods}
In recent years, a wide range of learning-based reconstruction methods has emerged. For a thorough overview of geometric deep learning, we refer readers to surveys such as ~\cite{Bronstein_2017, xiao2020survey}.
One early method, Deep Geometric Prior (DGP) ~\cite{williams2019deep}, divides the point cloud into patches and trains separate MLPs per patch, though it still depends on Poisson reconstruction for final surface extraction. Surface wrapping and iterative contraction play important roles in several learning-based pipelines. A notable example is Point2Mesh ~\cite{hanocka2020point2mesh}, which employs MeshCNN—an edge-based convolutional network with shared weights—to learn a self-prior and progressively refine an initial surface toward the target shape. While it shows promising reconstruction quality, the approach requires many iterations and substantial runtime (around three hours) to converge.
Deep learning has also been applied to implicit reconstruction of large-scale scenes. For instance, ~\cite{peng2020convolutional} uses a convolutional decoder to predict occupancy from local voxel crops, combining local and global cues for scalable scene reconstruction. Similarly, ~\cite{2020LocalImplicit} introduces Local Implicit Grid Representations, which exploit the prior that most surfaces exhibit geometric regularity at an intermediate scale, enabling robust reconstruction from partial or noisy inputs.
A number of studies ~\cite{2020Points2Surf, 2020Implicit, SIREN, wang2023neural, dong2024neurcadrecon} have focused on predicting signed neural distance fields. However, such methods often struggle to preserve fine geometric details and sharp features. Even with advanced network architectures, implicit representations based on signed distance functions tend to produce overly smooth surfaces and cannot faithfully capture true geometric discontinuities.

\subsection{Interpolation Methods}
Interpolation-based methods aim to preserve maximum fidelity from the original data. However, their performance tends to degrade when the input point cloud exhibits non-uniform distribution or significant noise. Existing interpolation approaches can be broadly categorized into two groups: traditional geometric methods and deep learning-based techniques.

\subsubsection{Traditional interpolation methods}
The fundamental prerequisite for interpolation methods is the ability to exactly interpolate all input points.
Existing techniques within this category can be broadly classified into four main approaches: tangent-plane methods, restricted Delaunay–based methods, sculpting algorithms, and region-growing strategies.
Tangent-plane approaches involve mapping points within a local neighborhood onto their tangent planes ~\cite{2010Surface} and inferring connectivity between them. However, constructing a watertight manifold remains challenging due to the difficulty of consistently stitching these local pieces.
Delaunay-based techniques, such as the well-known Crust ~\cite{amenta1998new} and Cocone ~\cite{amenta2000simple}, leverage Delaunay triangulation to generate high-quality triangles. While they perform well when the sampling is sufficiently dense relative to the local feature size, this condition is often hard to satisfy in practice.
Sculpting algorithms ~\cite{10.1016/j.cad.2014.08.021} take a volumetric approach by tetrahedralizing the convex hull and classifying interior versus exterior tetrahedra. Alternatively, ~\cite{Kuo_2005} proposed a greedy region-growing method based on the empty-ball principle, which struggles with sparse sampling, sharp features, and thin structures.
Region-growing methods include the Ball Pivoting Algorithm ~\cite{bernardini1999ball}, which rolls a ball of fixed radius to connect triplets of points into triangles. Its extension, the Intrinsic Property Driven (IPD) algorithm ~\cite{LIN20041}, improves point selection using a weighted least-length criterion. Further refinements were introduced by ~\cite{li2009surface}, which prioritizes growing in relatively flat regions before handling sharp features.
In summary, point insufficiency remains the most significant challenge for explicit interpolation approaches, as existing connection schemes often fail in complex geometric and topological settings.

\subsubsection{Learning-Based interpolation methods}
There exist several deep learning-based techniques for interpolating a point cloud. Among the latest advancements, Sharp and Ovsjanikov~\shortcite{sharp2020pointtrinet} introduced a notable method called \textit{PointTriNet}. This approach is unique in that it facilitates point set triangulation as an integral component within the 3D learning pipeline. PointTriNet comprises two neural networks: a classification network tasked with determining whether a given candidate triangle should be included in the triangulation, and a proposal network that generates additional triangle candidates. However, one limitation of PointTriNet is its limited ability to handle low-quality point clouds, resulting in output that may not constitute a manifold mesh.

\subsection{Restricted Delaunay Triangulation in Surface Reconstruction}
Restricted Delaunay Triangulation (RDT) has proven beneficial in mesh extraction~\cite{2014Automatic, 2016Fixed, 2017Surface, yan2009isotropic}, owing to its ability to generate high-quality triangulations directly from unstructured point sets. 
Initially, it was employed to enhance meshing quality~\cite{yan2009isotropic}.
In~\cite{2014Automatic}, the authors proposed remeshing the surfaces of 3D sealed geological structural models, aligning them with user-specified contact lines.
Meanwhile, ~\cite{2016Fixed} observed that if the input points, sampled from a given mesh surface, adhere to the local-feature-size (LFS) criterion, they can be connected into a triangle mesh, forming an alternative triangulation of the manifold. Nevertheless, this LFS-sampling condition often remains elusive in practical applications. 
Alternatively,~\cite{2017Surface} suggested treating each point as a disk oriented orthogonally to the estimated normal direction and subsequently connecting points based on the proximity of these disks to form a surface. 
However, this proxy surface is neither watertight nor manifold, and it contains significant overlapping regions.
As a result, this approach tends to yield numerous non-manifold edges.
To address this, ~\cite{wang2022restricted} proposed a progressive interpolation method based on RDT.
They start with a watertight manifold sphere as the initial guiding surface, then compute the RVD on the guiding surface, and finally update the guiding surface using RDT. 
The key challenge lies in ensuring the guiding surface remains a watertight manifold. 
~\cite{wang2022restricted} summarized manifold conditions and proposed a corresponding manifold fix strategy to guarantee that the guiding surface consistently stays watertight.
However, their method may fail when the shape represented by the point cloud contains deep internal cavities or thin plates.
Firstly, the guiding surface cannot be attracted to adhere to the internal cavity.
To approximate the real shape, they employed a heuristic sculpting approach that easily confuses the interior and exterior. 
Secondly, their algorithm struggles to converge because, in concave regions, each iteration may only interpolate a few points.
Finally, their manifold fix strategy only considered manifold conditions while overlooking the true topological structure, making it prone to confusing the two sides of thin-plate models.
To address these issues, we proposes a more robust and elegant algorithm based on RDT in this paper.

\section{Algorithm Overview}

\begin{figure*}[ht]
  \begin{center}
  \includegraphics[width=0.9\textwidth]{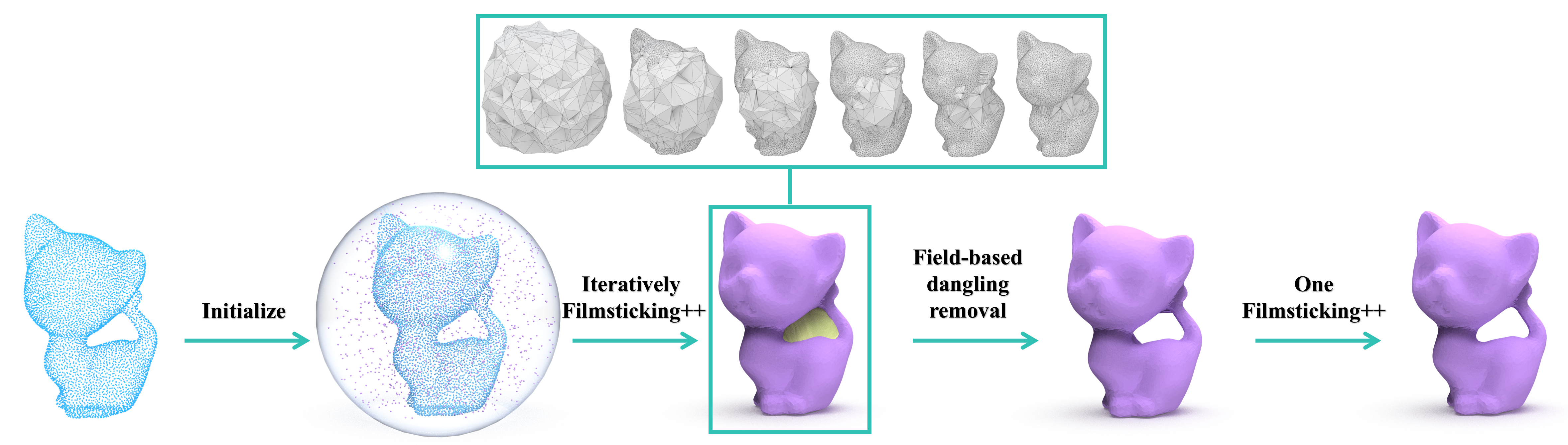}
    \makebox[0.196\textwidth][c]{\small (a) Input}
    \makebox[0.196\textwidth][c]{\small (b) Initialize }
    \makebox[0.196\textwidth][c]{\small (c) Filmsticking++}
    \makebox[0.196\textwidth][c]{\small (d) Dangling remove}
    \makebox[0.196\textwidth][c]{\small (e) Output}
  \vspace{-4mm}
  \end{center}
   \caption{
   The pipeline of our algorithm. (a) A point cloud as input. 
   (b) The bounding sphere as an initial guiding surface as well as a set of sampling points (typically around 1K) inside the guiding surface. 
   (c) An iterative process until the guiding surface interpolates nearly all points.
   (d) Field-based dangling removal for fixing the topology.
   (e) A single Filmsticking++ operation to accomplish the reconstruction.
   }
   \label{fig:Pipeline}
   \vspace{-2mm}
\end{figure*}

The goal of this paper is to generate a watertight and manifold triangle mesh $\mathbf{G} \in \mathbb{R}^3$ that faithfully captures the underlying shape. Note that the input point cloud $\mathbf{P} \in \mathbb{R}^3$ may have various imperfections, such as low or irregular point density, noise, missing data, and outliers, while the target surface is very geometrically and topologically complicated, featuring plates, tubes, sharp features, rich details, or even small topological holes.       

As illustrated in Fig.~\ref{fig:Pipeline}, our algorithm takes the watertight manifold bounding sphere as an initial guiding surface $\mathbf{G}$. 
At the same time, we initialize a set of virtual sites $V$ inside $\mathbf{G}$. 
These points facilitate the rapid evolution of $\mathbf{G}$, enabling it to naturally adapt to complex geometries such as internal cavities.
After that, we iteratively perform Filmsticking++ such that $\mathbf{G}$ gradually adheres to the input point cloud $\mathbf{P}$. 
This process terminates when $\mathbf{G}$ interpolates all points, excluding those that would result in a non-manifold geometry.
Finally, the entire reconstruction is completed with a field-based method to fix the topology, and a single Filmsticking++ operation to yield the reconstruction surface.
Note that, during the entire process, $\mathbf{G}$ is maintained to be a manifold.

\textbf{Filmsticking++} take input point cloud $\mathbf{P}$ and virtual sites $V$ as sites and compute the RPD on the guiding surface $\mathbf{G}$, and then update the virtual sites $V$ by delete attracted to $\mathbf{G}$ and $\mathbf{G}$ using RDT by dual of RPD, as shown in Fig.~\ref{fig:Pipeline}(c).
Details can be found in Section~\ref{Filmsticking}. 
It outperforms filmsticking~\cite{wang2022restricted} in three ways: 1) ensures that all points can be attracted to $\mathbf{G}$, 2) rapid evolution of $\mathbf{G}$, and 3) correctly distinguish the two sides of thin plates.

\textbf{Field-based dangling removal} is fix the topology based on field which estimated by guiding surface.
A straightforward and effective approach is to estimate normal vectors from the guiding surface, and then use the point cloud with normals to infer the inside/outside of the underlying shape. It is worth noting that our guiding surface closely adheres to almost all points and highly approximates the true shape, thus providing normals that are close to the ground truth. 
Multiple options are available, including winding number, Poisson reconstruction, and RBF-based interpolation.
Details can be found in Section~\ref{Sculpting}.

\section{Filmsticking++}\label{Filmsticking}

We begin by reviewing the concepts of the restricted Voronoi diagram (RVD), restricted power diagram (RPD), and restricted Delaunay triangulation (RDT). 
Next, we explain how RPD can be employed to attract the guiding surface. 
Following that, we outline the strategy for sampling virtual sites to accelerate the adherence process of the guiding surface. 
Finally, we introduce a smoothness-driven RPD fixing strategy that proves beneficial in the context of explicit surface reconstruction.

\subsection{RVD, RPD and RDT}
Let $\mathbf{P}=\{p_i\}$ be a collection of sites.
The Voronoi diagram of $\mathbf{P}$ is composed of $|\mathbf{P}|$ Voronoi cells, 
with each cell defined as follows:
\begin{displaymath}
C_i = \{x\in \mathrm{R}^3  \;\big|\; ||p_i - x|| \leq ||p_j - x||, \forall i \neq j\}.
\end{displaymath}

The power diagram can be treated as an extended Voronoi diagram. It allows each site $p_i$ to be equipped with a weight $w_i$. A power cell is defined as follows:
\begin{displaymath}
PC_i = \{x\in \mathrm{R}^3  \;\big|\; ||p_i - x||^2-w_i \leq ||p_j - x||^2-w_j, \forall i \neq j\}.
\end{displaymath}

Given a two-manifold surface $S \in R^3$, and a set of finite sites $\mathbf{P}$, the RVD is defined as the intersection of the 3D Voronoi diagram of $\mathbf{P}$ and the surface $S$.
Similarly, the RPD is defined as the intersection of the 3D Power diagram of $\mathbf{P}$ and the surface $S$. 
The dual of the RVD or RPD, called the RDT.

\subsection{RPD on the Guiding Surface}~\label{RPD_GuidingSurface}

\begin{figure}[!t]
  \begin{center}
  \includegraphics[width=0.95\columnwidth]{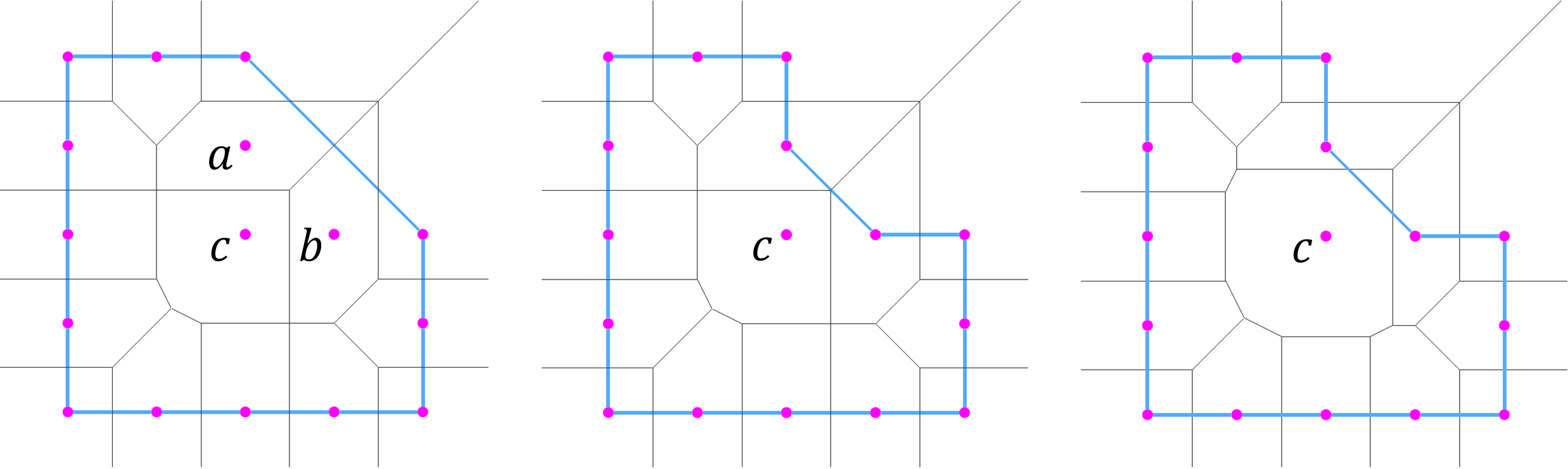}
  \makebox[0.15\textwidth][c]{\small (a) RVD}
  \makebox[0.15\textwidth][c]{\small (b) RVD}
  \makebox[0.15\textwidth][c]{\small (c) RPD}
  \vspace{-4mm}
  \end{center}
   \caption{By replacing RVD with RPD, we can attract point that far away from guiding surface to the guiding surface.
   (a) Point $a$, $b$ is near the guiding surface, thus dominating an area of guiding surface.
   (b) With RVD, point $a$, $b$ is embedded into the dual triangulation, but dominated region of point $c$ has no intersection with the guiding surface, the point $c$ cannot be attracted onto the guiding surface.
   (c) With RPD, the weight of $c$ is increased so that it can be attracted to the surface successfully.
   In the case of RVD, only those points near the guiding surface can dominate an area. In this paper, we extend RVD to RPD, allowing distant points to adhere to the guiding surface by increasing their weights.}
   \label{fig:RVD_vs_RPD}
   \vspace{-2mm}
\end{figure}

Different from conventional application scenarios where all the sites are situated on the surface, in explicit surface reconstruction, the underlying surface is unknown until the reconstruction phase concludes. 
Therefore, in~\cite{wang2022restricted}, the guiding surface $\mathbf{G}$ is introduced to facilitate a surface evolution process. This allows the guiding surface to iteratively adhere to the point cloud. 
During the iterations, points are not necessarily located on the surface.

The key concept behind filmsticking in~\cite{wang2022restricted} lies in the observation that points near the guiding surface will be attracted onto the surface. 
See Fig.~\ref{fig:RVD_vs_RPD}(a), point $a, b$ is near the guiding surface, thus dominating an area of $\mathbf{G}$. 
Based on the proximity within the restricted Voronoi diagram, point $a, b$ is embedded into the dual triangulation in Fig.~\ref{fig:RVD_vs_RPD}(b). 
On the other hand, point $c$ is too distant from $\mathbf{G}$, and as a result, $c$'s Voronoi cell does not intersect with $\mathbf{G}$. Therefore, point $c$ is not included in the dual triangulation.

By considering the surface decomposition from the power diagram, we increase the weight of $c$ from 0 to a certain value, as shown in Fig.\ref{fig:RVD_vs_RPD}(c). It can be seen that $c$'s cell is enlarged, thereby dominating an area of $\mathbf{G}$. 
This is different from the situation in the Voronoi diagram. 
This insight inspires us to employ RPD as a tool to expedite the attraction of the guiding surface. In the following, we will discuss how to set the weight.

Let the input point cloud be $\mathbf{P} = \{p_i\}$, the guiding surface is $\mathbf{G}$ that its vertex set is $\mathbf{V}=\{v_i\}$, 
the points not lying on $\mathbf{G}$ define as $\mathbf{Q} = \{q_i\}$, $\mathbf{Q} \cup \mathbf{V} = \mathbf{P}$.

\paragraph{Case 1}
\begin{wrapfigure}{r}{2cm}
\vspace{-0.5mm}
  \hspace*{-4mm}
  \centerline{
  \includegraphics[width=25mm]{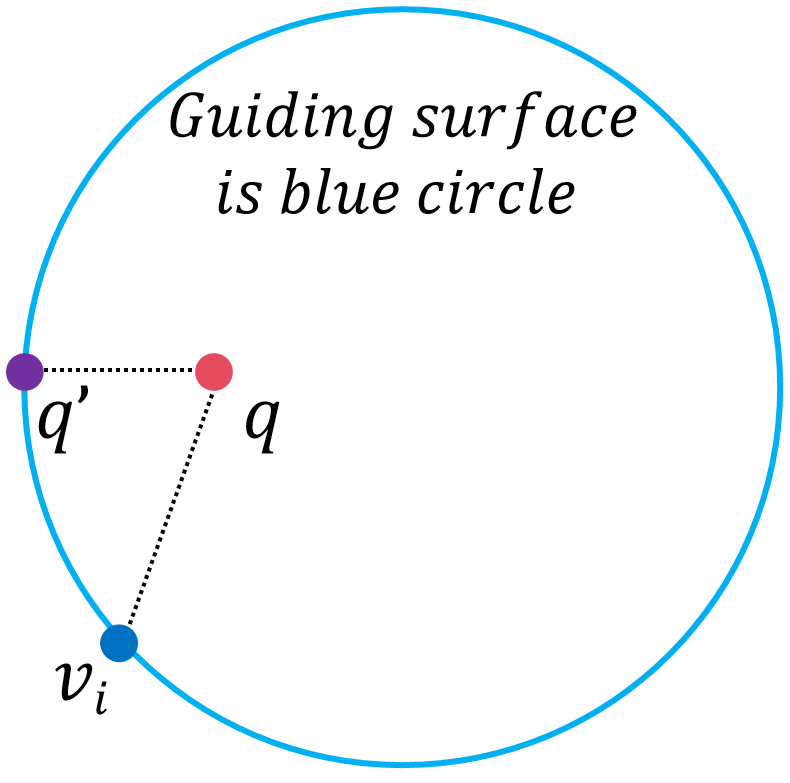}
  }
  \vspace*{-4mm}
\end{wrapfigure}
If number of set $\mathbf{Q}$ is 1, only one site $q$ has not been attracted onto $\mathbf{G}$, i.e., $q \notin \mathbf{V}$. 
The projection of $q$ onto $\mathbf{G}$ is denoted by $q'$, and the squared distance from $q$ to $\mathbf{G}$ is $\|q - q'\|^2$.
For all sites $v_i \in \mathbf{V}$ that have already been attracted to $\mathbf{G}$, their weights are set to zero: $W_{v_i} = 0$.
If the weight $W_q$ of site $q$ satisfies $W_q > \|q - q'\|^2$, then for any site $v_i \in \mathbf{V}$, the projection point $q'$ must satisfy:
\begin{equation}
\|q - q'\|^2 - W_q  < \|v_i - q'\|^2 - W_{v_i}.
\end{equation}
where $W_{v_i} = 0$, $q$ must dominates a local region of $\mathbf{G}$ near $q'$, and the power cell of $q$ intersects $\mathbf{G}$.
Additionally, to ensure that the power cell of point $q$ does not encroach any site $v_i$ lying on $\mathbf{G}$, the weight of $q$ must satisfy:
\begin{equation}
W_{q} < \min_{v_i \in \mathbf{V}} \|q - v_i\|^2.
\end{equation}
Therefore, a site $q$ can be attracted onto $\mathbf{G}$ without altering the local topology of the power diagram if its weight $W_{q}$ satisfies:
\begin{equation}\label{eq:weight_condition}
\|q-q'\|^2 < W_q \;<\; \min_{v_i \in \mathbf{V}} \|q - v_i\|^2.
\end{equation}
Note that such a $W_q$ may not always exist. 
If $q'$ coincides with a vertex $v_i$ of $\mathbf{G}$, $\|q-q'\|^2 = \|q - v_i\|^2$, then no weight exists that can guarantee the $q$ attracted onto the guiding surface $\mathbf{G}$.
When this occurs, we apply a slight perturbation to $q$. In practice, however, the probability of encountering such a situation is extremely low.

\paragraph{Case 2}
\begin{wrapfigure}{r}{2cm}
\vspace{-0.5mm}
  \hspace*{-4mm}
  \centerline{
  \includegraphics[width=25mm]{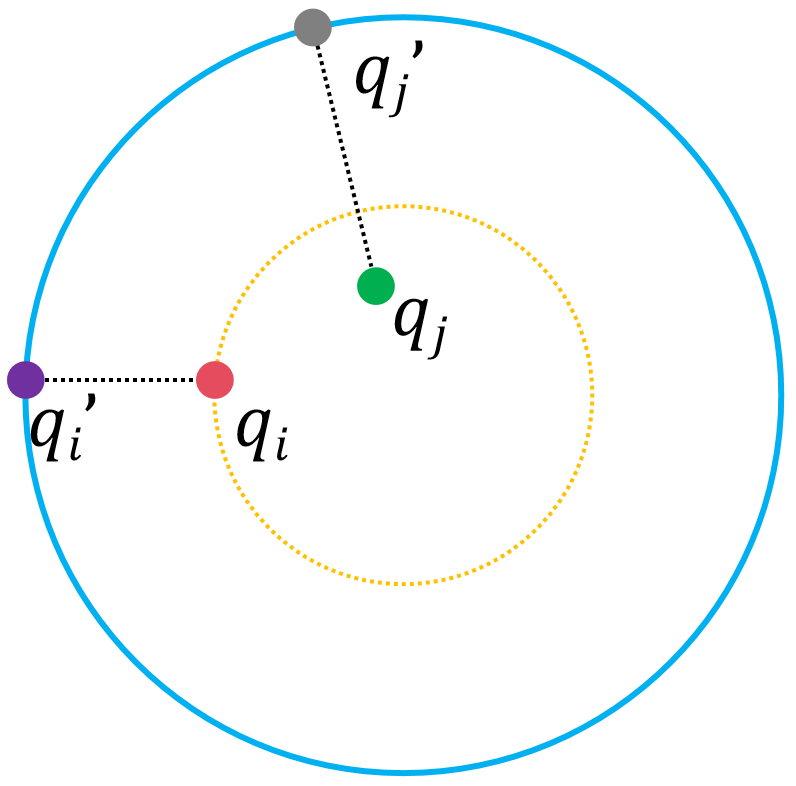}
  }
  \vspace*{-4mm}
\end{wrapfigure}
If number of set $\mathbf{Q}$ more than 1, there are more than one point that have not been attracted onto the guiding surface $\mathbf{G}$.
Assume that the projections onto $\mathbf{G}$ of all points in $\mathbf{Q}$: 1) do not coincide with any vertex in $\mathbf{V}$ of $\mathbf{G}$, and 2) are mutually non-coincident.
If each site $q_i \in \mathbf{Q}$ is assigned the weight $W_{q_i} = \|q_i - q_i'\|^2$, $q_i'$ is projection of $q_i$ on $\mathbf{G}$, then there must exist a point $q_i \in \mathbf{Q}$ that can be attracted onto the guiding surface.
For any $q_i \in \mathbf{Q}$, satisfy $\|q_i - q_i'\|^2 - W_{q_i} = 0$.
Let $q_i \in \mathbf{Q}$ be the site closest to $\mathbf{G}$. 
Recalling our assumption, the projections of $\mathbf{Q}$ onto $\mathbf{G}$ are mutually non-coincident ($q_i' \neq q_j'$ when $j \neq i$), then satisfy $\|q_j - q_j'\|^2 - W_j  < \|q_j - q_i'\|^2 - W_j$, then must satisfy:
\begin{equation}
\|q_i - q_i'\|^2 -W_{q_i} < \|q_j - q_i'\|^2 - W_{q_j}, \forall q_j \in \mathbf{Q},\; j \neq i.
\end{equation}
It is evident that $q_i$ can be successfully attracted onto $\mathbf{G}$.
Note that if the second assumption is not valid, $q_j' = q_i'$ (i.e., their projections coincide), $\|q_j - q_i'\|^2 - W_{q_j} = 0$, making the attribution of $q_i'$ ambiguous in the power diagram.
To resolve this, we reduce the weight $W_{q_j}$ of each $q_j \in \mathbf{Q} \;(j \neq i)$.
This ensures that the region near $q_i'$ is strictly dominated by $q_i$, allowing $q_i$ to be attracted onto $\mathbf{G}$.

Finally, for all sites not residing on $\mathbf{G}$, we set their weights to $\|q_i - q_i'\|^2$. 
It ensure that at least one site can be successfully attracted onto $\mathbf{G}$. 
Although this option is conservative, it is superior to RVD in that the filmsticking process cannot terminate until nearly all points are attracted onto $\mathbf{G}$.

\begin{figure}[!t]
  \begin{center}
  \includegraphics[width=0.95\columnwidth]{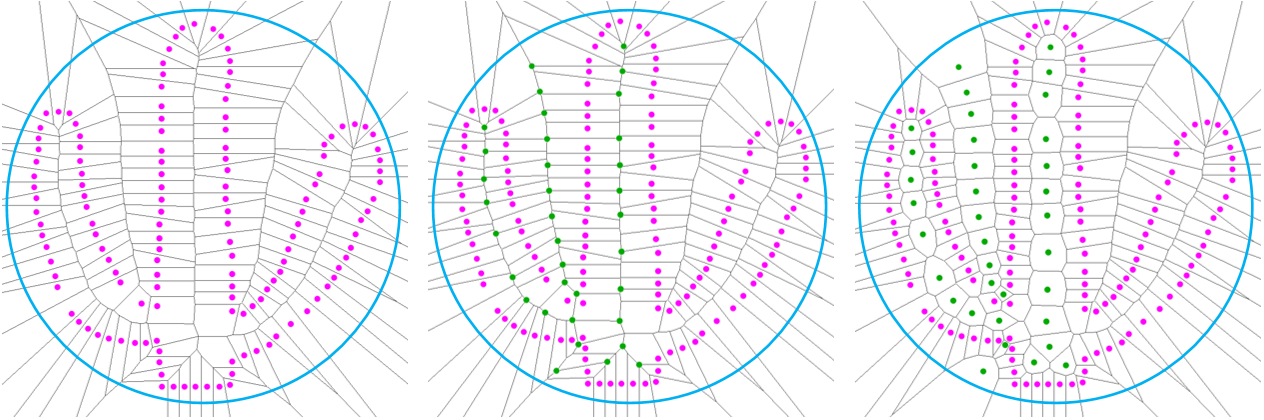}
  \makebox[0.155\textwidth][c]{\small (a) Input }
     \makebox[0.155\textwidth][c]{\small (b) Virtual sites }
      \makebox[0.155\textwidth][c]{\small (c) Update Voronoi}
  \includegraphics[width=0.95\columnwidth]{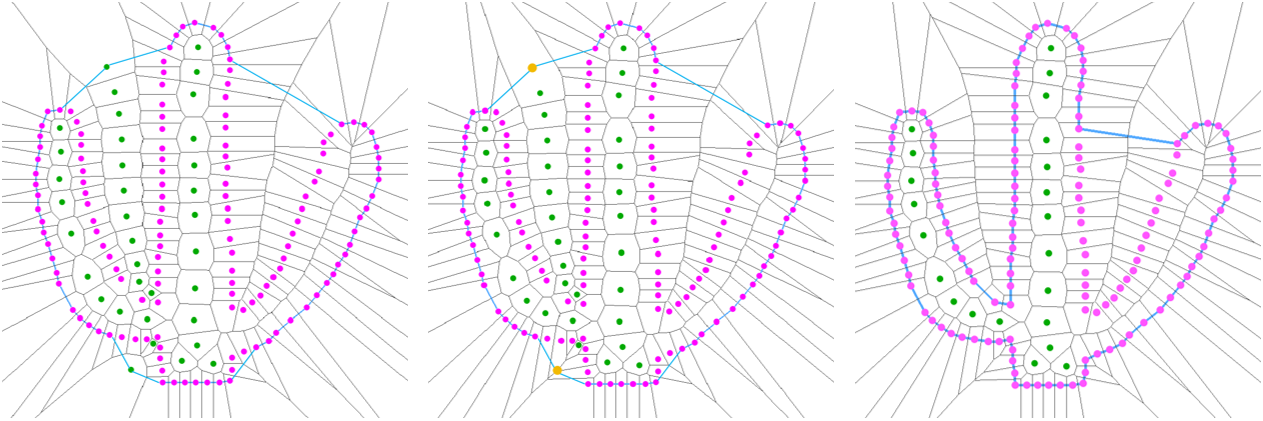}
  \makebox[0.155\textwidth][c]{\small (d) One Filmsticking++}
     \makebox[0.155\textwidth][c]{\small (e) Remove virtual sites}
      \makebox[0.155\textwidth][c]{\small (f) Nine Filmsticking++}
  \vspace{-4mm}
  \end{center}
    \caption{Virtual vertices to accelerate the evolution of guiding surface. 
    (a) Input point cloud. 
    (b) Sample a set of virtual sites (colored in green) inside the guiding surface, to test role of virtual points, deep internal cavities are set up with and without virtual sites, respectively.
    (c) Update Voronoi diagram including virtual sites.
    (d) One Filmsticking++, a virtual site is attracted onto the guiding surface, and update the guiding surface.
    (e) The virtual site (yellow) is removed, allowing the guiding surface to be re-partitioned by the neighboring sites.
    (f) After 9 Filmsticking++, guiding surface attracted all points in deep internal cavities with virtual sites.
   }
   \label{fig:virual_sites}
   \vspace{-2mm}
\end{figure}

\subsection{Virtual Sites}~\label{VirtualSites}
Although the RPD can ensure all sites are attract to guiding surface, it may still struggle when dealing with highly curved underlying shapes.
In the worst case, only one site will be attracted to guiding surface at one iteration.
In the following, we introduce a straightforward yet effective sampling technique to accelerate the evolution of guiding surface.

Recall that the initialization tasks are two-fold: introducing a bounding sphere as the guiding surface $\mathbf{G}$ and sampling a set of virtual sites inside this sphere. 
See Fig.~\ref{fig:virual_sites} for an illustration. 
The virtual sites are allowed to be attracted onto the guiding surface (Fig.~\ref{fig:virual_sites}(d)).
However, once a virtual site is located on the guiding surface, it is removed and cannot be attracted again, and leaving their dominating regions to be re-partitioned by nearby sites (Fig.~\ref{fig:virual_sites}(e)).
Once all sites are attracted onto the guiding surface, all remaining virtual sites will be removed.
In the following, we will discuss how to set the location of virtual sites.

We observe that as evolution of guiding surface, the external medial axis of the real shape is progressively excluded outside the guiding surface. 
Therefore, quickly excluding the external medial axis to the exterior of the guiding surface is equivalent to accelerating the evolution of the guiding surface.
Based on it, we select virtual sites along the medial axis direction, as shown in Fig.~\ref{fig:virual_sites}(b), aligning the direction of greatest change in the guiding surface with the external medial axis to accelerate convergence. 

In practice, we cannot extract the exact external medial axis from the point cloud. 
Therefore, we approximate the medial axis using Voronoi vertices and treat them as virtual sites. 
This approach introduces two issues: 1) virtual points are also placed on unwanted inner medial axis, and 2) when the point cloud is large, the number of Voronoi vertices becomes excessively high.

Regarding the first issue, virtual points on the inner medial axis are typically surrounded by the point cloud, and their Voronoi cells are obstructed by the point cloud, preventing them from being easily attracted onto the guiding surface, see Fig.~\ref{fig:virual_sites}. Even if a virtual point on the inner medial axis is incorrectly attracted onto the guiding surface due to missing data in the point cloud, it will be removed in the next iteration, and the guiding surface will revert to its original state.

Regarding the second issue, an excessive number of virtual points clearly slows down convergence. However, too few virtual points fail to provide effective acceleration. Consider a scenario with an infinitely dense point cloud describing a deeply concave shape: if exactly one virtual point is attracted onto the guiding surface in each iteration, the process terminates once all virtual points on the outer medial axis have been removed. Thus, the number of virtual points directly determines the iteration count. 
However, determining the minimal number of virtual points while ensuring at least one is attracted to guiding surface per iteration is a theoretically challenging and practically infeasible problem.
To address this, we employ a heuristic strategy: perform farthest point sampling on the Voronoi vertices to achieve a relatively uniform distribution. Through experiments, we find that sampling approximately 1k points is an ideal choice. 
This approach significantly accelerates the evolution of the guiding surface and keeps the iteration count within 20 for nearly all models, as shown in Fig.~\ref{fig:virual_sites}(f).

\subsection{Smoothness Based Manifoldness Fixing}
\paragraph{Manifoldness conditions}
Suppose that $p_1,p_2,p_3\in \mathbf{P}$ commonly determine a RVD vertex, then $p_1,p_2,p_3$ are connected into a triangle in the dual structure RDT.
The conditions for enforcing RDT to be a manifold if the following conditions are satisfied at the same time~\cite{1997Triangulating,STEPHAN2005Automatic,wang2022restricted}:
 \begin{enumerate}
    \item Any cell of RVD must be homeomorphic to a 2D disk;
    \item Any two neighboring cells of RVD must have one common edge (rather than multiple segments).
    \item Any cell of RVD has at least three neighbors.
\end{enumerate}

\begin{wrapfigure}{r}{2cm}
\vspace{-0.5mm}
  \hspace*{-4mm}
  \centerline{
  \includegraphics[width=25mm]{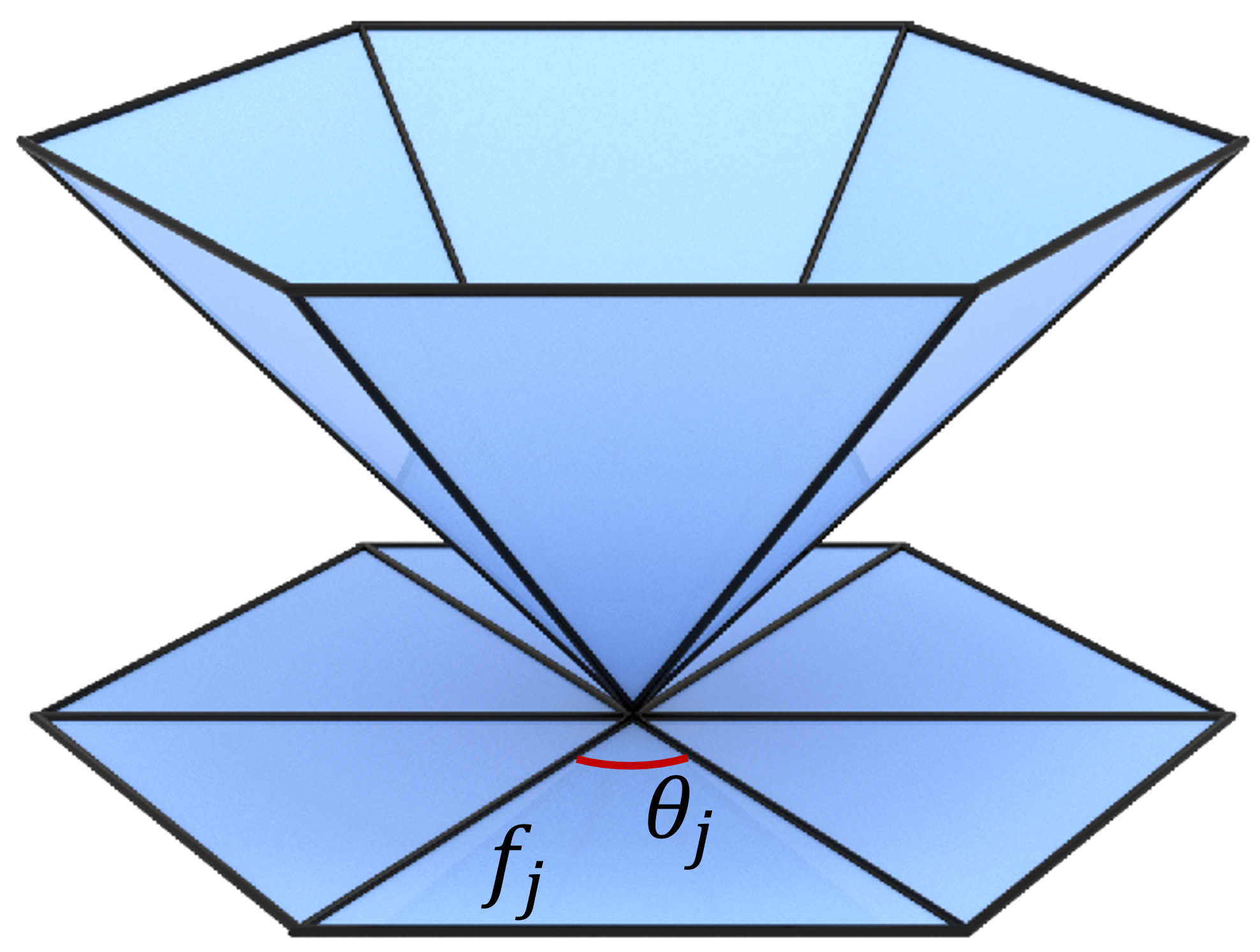}
  }
  \vspace*{-4mm}
\end{wrapfigure}
\begin{figure}[!ht]
  \begin{center}
  \includegraphics[width=0.95\columnwidth]{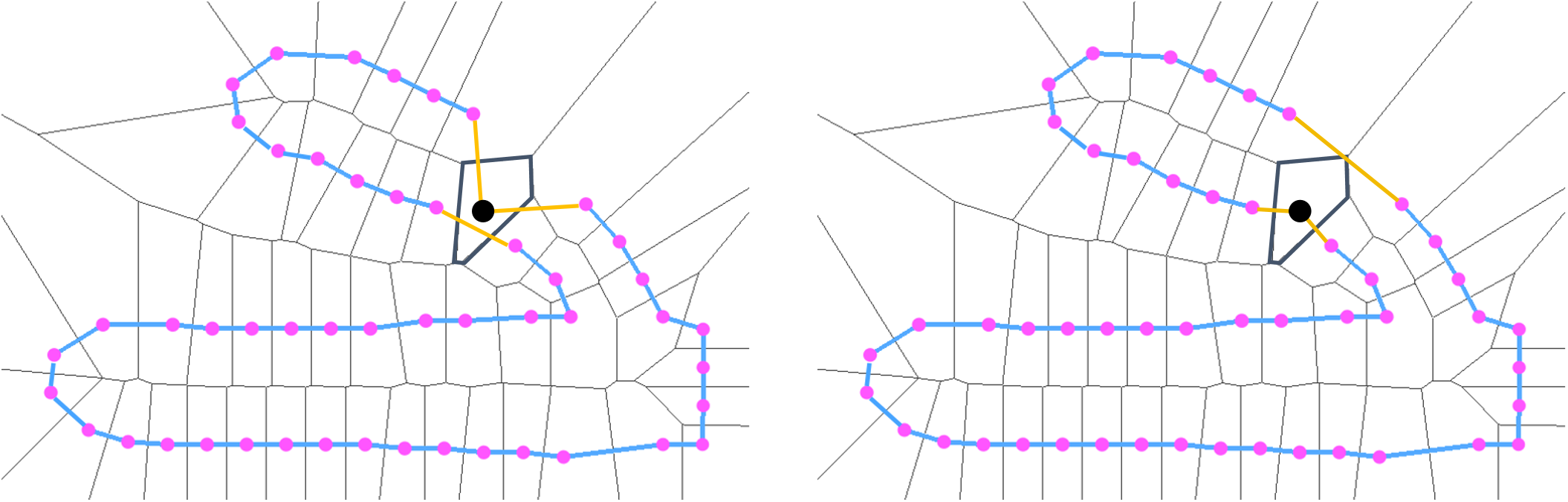}
  \makebox[0.225\textwidth][c]{\small (a) Incorrect topology}
      \makebox[0.225\textwidth][c]{\small (b) Correct topology}
  \vspace{-4mm}
  \end{center}
   \caption{Smoothness manifold fix. When a site (black) dominates two separate regions of the guiding surface, resulting in RDT a non-manifold vertex. (a) Filmsticking~\cite{wang2022restricted} determines the dominated region of a site based on the distances from the site to its multiple disconnected dominated regions, it may cause many unwanted topological.
   (b) We propose a smoothness-driven principle to select one of the regions and associate the site with the selected region.
   }
   \label{fig:SmoothManifoldFix}
   \vspace{-2mm}
\end{figure}

Any violation of the manifoldness conditions falls within one of the four situations; 
See~\cite{wang2022restricted} for detailed discussion. 
It must be noted that the manifoldness conditions still hold for RPDs. 
Imagine that the input is a thin-plate point cloud, the most common occurrence is when there is a site dominating two separate regions, see Fig.~\ref{fig:SmoothManifoldFix} (a), causing two umbrellas to be adjoined tip to tip. 
In this case, ~\cite{wang2022restricted} determines the dominated region of a site based on the distances from the site to its multiple disconnected dominated regions, it may cause many unwanted topological while trying to satisfy the manifoldness conditions.

However, following the simplicity principle, the simpler scenario is preferred when ambiguity arises. In other words, when the site~$v_i$ dominates two separate regions, we prefer the choice that enables the guiding surface to have a simple and smooth shape. 
For this purpose, we first separate the triangles incident to~$v_i$ 
into rings based on the adjacency-based dual graph,
and then for each ring of triangles, we evaluate the smoothness degree using the following formula (refer to the inset for an illustration):
\begin{equation}
\xi(v_i) = \sum_{f_j}\theta_j \lVert \mathbf{n}_j - \overline{\mathbf{n}_i} \rVert^2,
\end{equation}
where $\overline{\mathbf{n}_i}$ is the average normal at $v_i$, $f_j$ is a triangle incident to $v_i$, $\mathbf{n}_j$ is the normal vector at $f_j$, and $\theta_j$ is the $v_i$-incident angle in $f_j$.
Based on this criterion, we have restored the correct topology in Fig.~\ref{fig:SmoothManifoldFix} (b).

\section{Field-based Dangling Removal}\label{Sculpting}

\begin{figure}[!ht]
  \begin{center}
  \includegraphics[width=0.95\columnwidth]{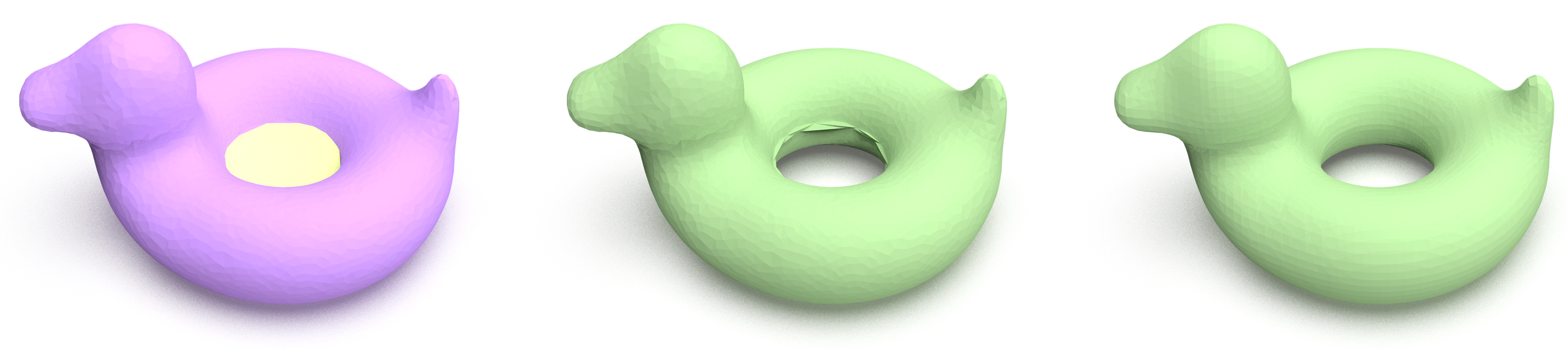}
  \makebox[0.15\textwidth][c]{\small (a) Topological plates}
      \makebox[0.15\textwidth][c]{\small (b) Sculpting}
      \makebox[0.15\textwidth][c]{\small (c) Poisson Reconstruction}
  \vspace{-4mm}
  \end{center}
   \caption{ Removal of unwanted topological plates.
   (a) Guiding surface contains two-layer ``plates'' with a narrow gap at the end of Filmsticking++.
   (b) We remove unwanted topological using Sculpting~\cite{wang2022restricted}. 
   (c) We using Poisson reconstruction to regenerate a correct topology.
   }
   \label{fig:dangling}
   \vspace{-2mm}
\end{figure}

To this end, the guiding surface~$\mathbf{G}$ adheres to the point cloud, with its topology possibly different from the underlying topology, see Fig.~\ref{fig:dangling}(a). 
In order to effectively detect and fix the topology, \citet{wang2022restricted} used the winding number~\cite{barill2018fast} to perform the sculpting operation. Their method begins with building tetrahedralization for the interior of~$\mathbf{G}$ with tetgen~\cite{10.1145/2629697}, obtaining a collection of tetrahedra $\{T_i\}_{i=1}^m$. The winding number of each $T_i\in \mathbf{T}$ can be evaluated as
\begin{equation}
 \label{eqn:01}
 W(T_i)=\sum_{j=1}^n a_j \frac{(p_j-t_i)\cdot \mathbf{n}_j}{4\pi \lVert(p_j-t_i)\rVert^3},
\end{equation}
where $t_i$ is the center point of $T_i$, $\mathbf{n}_j$ is the normal vector at $p_j$, $n$ is the total number of vertices of $\mathbf{G}$, and $a_j$ is the influence area of $p_j$, computed by projecting the neighboring points onto the best-fit plane and querying the area of its 2D Voronoi cell.

However, our resulting guiding surface $\mathbf{G}$ tightly wraps the given point cloud. 
For where it differs from the underlying surface, $\mathbf{G}$ contains two-layer ``plates'' with a narrow gap, which depends on point density, see Fig.~\ref{fig:dangling}(a).
Subsequently, we can perform sculpting or implicit reconstruction based on the guiding surface, with the key advantage that accurate normal vectors can be extracted from the guiding surface generated by Filmsticking++. 
Using our guiding surface, we tested both sculpting~\cite{wang2022restricted} and Poisson reconstruction, obtaining nearly identical results, as shown in Fig.~\ref{fig:dangling}.
In this paper, Poisson reconstruction is adopted by default to remove unwanted topological plates.

\section{Experiments}\label{Experiments}
In this section, the details regarding experiment setting are first presented, 
followed by an explanation of the metrics and evaluation protocol.
we evaluate the proposed approach on various datasets, including both synthetic and real scanned point clouds. 
The experiments were all conducted on a PC which have an Intel(R) Core i9-13900k CPU 3.60 GHz with 64 GB memory and an NVIDIA Geforce RTX 3090 graphics card. 

\subsection{Experimental Setting}
\paragraph{Existing Approaches.} 
To demonstrate the superiority of our approach, we conduct a comparison with eight surface reconstruction methods.
Among these, some are interpolation-based, including Greedy~\cite{10.1007/s00371-003-0217-z}, Ball Pivoting (BP)~\cite{bernardini1999ball}, and the recently introduced Restricted Delaunay (RD)~\cite{wang2022restricted}.
Others are approximation-based, such as Screened Poisson (SPR)~\cite{kazhdan2013screened}, Neural-Singular-Hessian (NSH)~\cite{wang2023neural}, NeurCADRecon~\cite{dong2024neurcadrecon} and NN-VIPSS~\cite{ju2025variational}.
Additionally, we include PointTriNet (PTN)~\cite{sharp2020pointtrinet} in our comparison, a learning-based method that interpolates the given point cloud.
It is noteworthy that Screened Poisson requires normal vectors for the point cloud, and we utilize the method proposed in~\cite{xu2023globally} for estimating these normals to facilitate the execution of Screened Poisson.

\paragraph{Datasets.}
We employed a range of datasets for our testing purposes.
The initial set of data was derived from direct sampling of meshes from Thingi10K~\cite{2019thingi10k} dataset.
Furthermore, we incorporated real-world scan data, which encompassed datasets such as~\cite{huang2022surface}, the Statue Model dateset~\cite{EPFL}, as well as the extensive dataset from~\cite{aanaes2016large} and ~\cite{wang2022restricted}.
Additionally, we used the SHINING 3D Einscan SE scanner to obtain 4 real scans.

\begin{figure*}[ht]
  \begin{center}
  \includegraphics[width=0.95\textwidth]{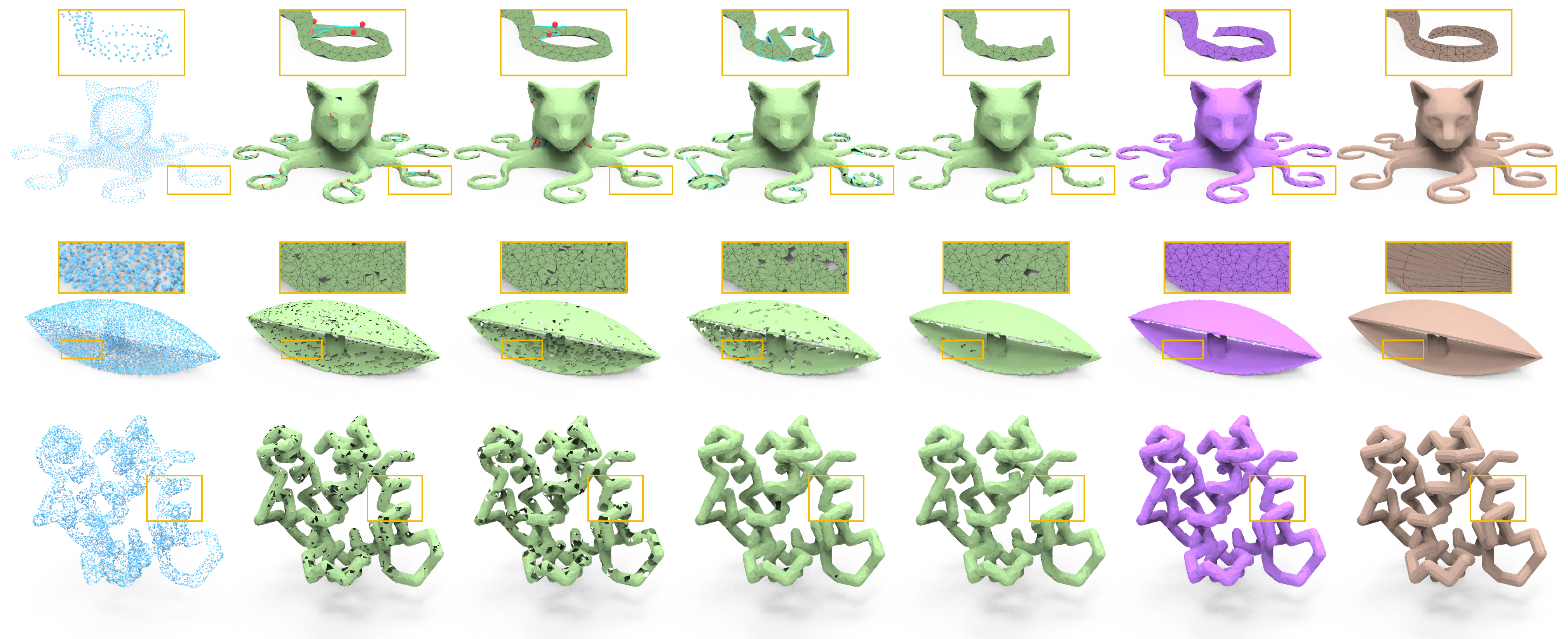}
  \makebox[0.13\textwidth][c]{\small (a) Input}
    \makebox[0.13\textwidth][c]{\small (b) PTN}
    \makebox[0.13\textwidth][c]{\small (c) BP}
    \makebox[0.13\textwidth][c]{\small (d) Greedy}
    \makebox[0.13\textwidth][c]{\small (e) RD}
    \makebox[0.13\textwidth][c]{\small (f) Ours}
    \makebox[0.13\textwidth][c]{\small (g) GT}
  \vspace{-4mm}
  \end{center}
   \caption{Comparison with explicit approaches. 
   Input point clouds are sampled on thingi10k using the Monte-Carlo, is 4k, 8k and 8k points from the "Fox", "Bifolium hyperboloid" and thin tube models "Polymer Chain".
   PTN and BP has a large number of non-manifold points/edges, resulted in numerous holes and incorrect topological connections.
   Greedy cannot guarantee a closed mesh, especially when dealing with low-quality point clouds.
   RD uses winding number to infer the potential interior and exterior, resulting in misjudged the interior and exterior of the potential shape, excessive sculpting, such as "Fox" and "Polymer Chain".
   Additionally, RD fail to distinguish the two sides of a thin plate, see "Bifolium hyperboloid".
   Our method, in contrast, can naturally adhere to the point cloud, leading to a more faithful reconstruction result. }
   \label{fig:compare_explicit}
   \vspace{-2mm}
\end{figure*}

\subsection{Comparison with Explicit Reconstruction Approaches}
As shown in Fig.\ref{fig:compare_explicit}, we respectively sampled 4k, 8k and 8k points from the "Fox", "Bifolium hyperboloid" and thin tube models "Polymer Chain" on thingi10k using the Monte-Carlo sampling, to assess the accuracy of the reconstructed mesh topology links. 
\paragraph{Comparison with RD~\cite{wang2022restricted}}
In this chapter, we will make a detailed comparison with it, and verify that our method is superior to RD.
RD is one of the latest explicit reconstruction algorithms. 
We specifically address the pain points of the RD method. 
RD uses filmsticking to interpolate the interior point cloud, but it requires a winding number to infer the potential interior and exterior of the model, resulting in misjudged the interior and exterior of the potential shape, excessive sculpting, see Fig.~\ref{fig:compare_explicit} "Fox". 
Secondly, once a point is interpolated onto the guiding surface, the connection relationship of this point will not be changed, even if it is incorrect. 
For example, RD will incorrectly connect the inner wall and outer wall of "Bifolium hyperboloid" model in Fig.~\ref{fig:compare_explicit}.
In comparison, we use smoothness manifold fix to help us achieve correct connect. 
Similarly, the RD result of the thin tube model shown in Fig.~\ref{fig:compare_explicit} is broken, mainly because most of the point clouds are not interpolated to the guiding surface when the phase of filmsticking ends, resulting in excessive dependence on the sculpting based on the winding number, which makes it impossible to accurately distinguish the inside and outside of the potential shape.

\paragraph{Manifoldness}
The requirement of ensuring a manifold and watertight surface may seem inconsequential for approximation-based techniques that simply extract the zero-level set of a scalar field.
However, for interpolation-based methods, this requirement becomes an exceedingly challenging task due to the inherent and often overwhelming combinatorial complexity.
Among all interpolation-based approaches, only our method and RD~\cite{wang2022restricted} can confidently guarantee manifoldness.
However, our method requires significantly fewer iterations compared to the RD method.
When considering the "Fox" model in Fig.~\ref{fig:compare_explicit}, PTN generates 183 non-manifold edges, 19 non-manifold vertices, and 132 open-boundary edges. BP, on the other hand, produces 73 non-manifold vertices and 258 open-boundary edges. Greedy, which relies on a greedy triangle selection strategy to predict the next optimal connection, yields 136 open-boundary edges. It's important to note that Greedy's best guesses are solely based on the local positional configuration of points, lacking any understanding of the overall shape. As a result, Greedy cannot guarantee a closed mesh, especially when dealing with low-quality point clouds that deviate significantly from the ideal Local Feature Size~(LFS) sampling conditions.

\paragraph{Topological Correctness}
We assess the accuracy of the reconstructed mesh topology links. 
We found that PTN, BP, and Greedy methods resulted in numerous holes and incorrect topological connections, as shown in Fig.~\ref{fig:compare_explicit}.
While RD maintained a watertight manifold, it unexpectedly created holes at the "Bifolium hyperboloid" model and suffered from breaking at various points along the "Fox" legs and "Polymer Chain" tube, ultimately failing to reconstruct a proper topology. 

\begin{figure*}[ht]
  \begin{center}
  \includegraphics[width=0.95\textwidth]{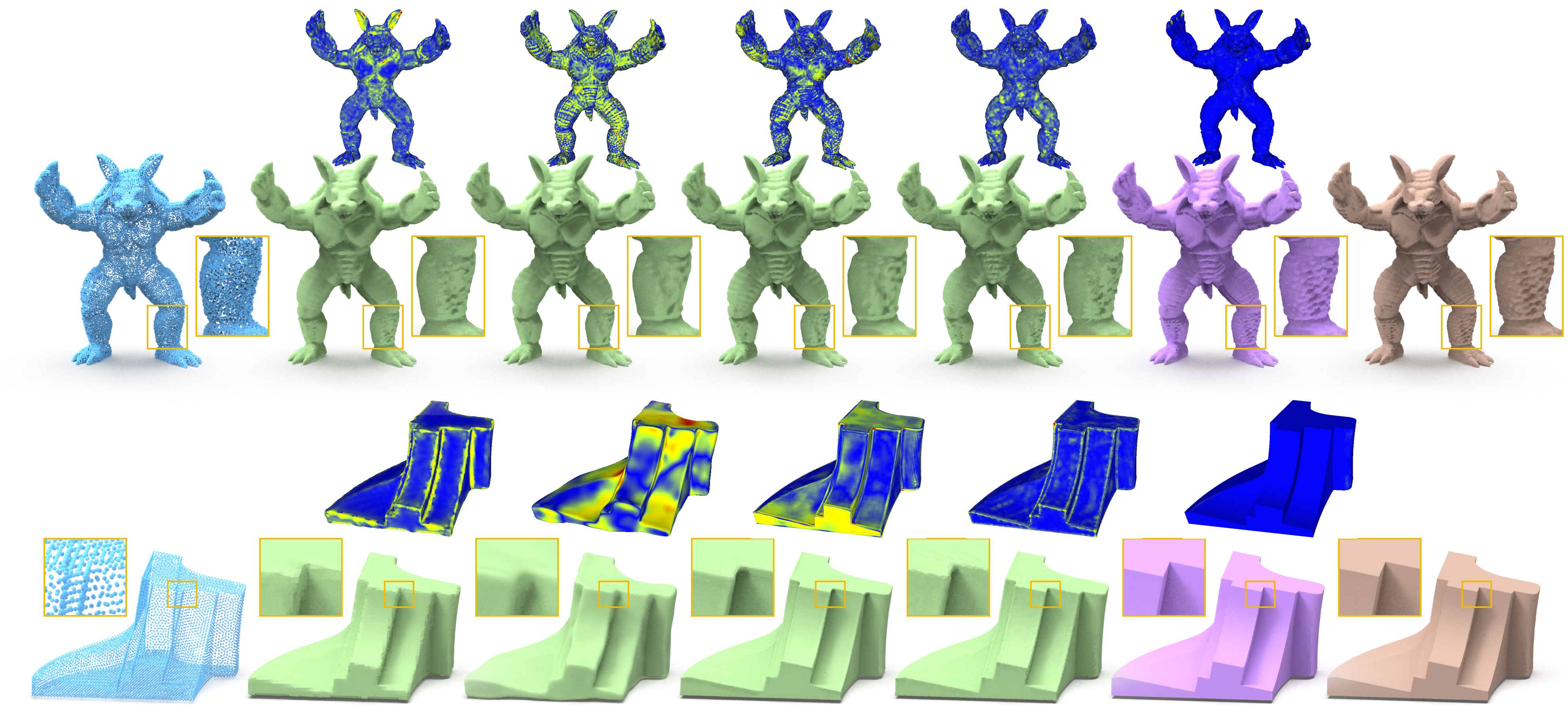}
  \makebox[0.13\textwidth][c]{\small (a) Input}
    \makebox[0.13\textwidth][c]{\small (b) SPR}
    \makebox[0.13\textwidth][c]{\small (c) NSH}
    \makebox[0.13\textwidth][c]{\small (d) NeurCAD}
    \makebox[0.13\textwidth][c]{\small (e) NN-VIPSS}
    \makebox[0.13\textwidth][c]{\small (f) Ours}
    \makebox[0.13\textwidth][c]{\small (g) GT}
  \vspace{-4mm}
  \end{center}
   \caption{Comparison with implicit approaches. The input points clouds accurately coincide with the surface of the model.
   Our approache, falling within the category of explicit reconstruction approaches, can preserve geometric details to the greatest extent, but implicit reconstruction algorithms cannot.
   See the highlighted windows. 
   By taking the vertices of a CAD-type polygonal model as input, our algorithm can precisely restore the original model,
   with a step of feature-aligned edge flip~\cite{wang2022restricted} as postprocessing.}
   \label{fig:compare_implicit}
   \vspace{-2mm}
\end{figure*}

\subsection{Comparison with Implicit Approaches}
We compare implicit methods in Fig.~\ref{fig:compare_implicit}.
In specific scenarios where the sampling points can accurately coincide with the surface of the model, interpolation methods emerge as distinctly superior to approximate techniques. Moving forward, we aim to showcase the unparalleled excellence of our proposed methodology in preserving features and capturing intricate geometric details.
\paragraph{Feature Lines}
We directly obtain mesh vertices (10K) from the ground truth to compare the implicit method, including Poisson reconstruction and the latest methods NSH. 
As shown in the Fig.~\ref{fig:compare_implicit} model "Fandisk", our method is almost the same as that of ground truth. 
This is also an essential reason for explicit reconstruction methods. 

\paragraph{Geometry Details}
The approximate method bears a stronger resemblance to smoothing, leading to a reconstructed surface that falls short of accurately interpolating a given point cloud. 
On the other hand, it becomes apparent that interpolation-based reconstruction excels in faithfully reproducing the defining features of the original shape; See Fig.~\ref{fig:compare_implicit} "Armadillo".

\begin{table*}[!t]
\centering
\caption{ Quantitative comparison on the Thingi10k dataset~\cite{2019thingi10k}. 
Each point cloud has 10K or 5K points. Within each column, the best scores are highlighted in bold.}
\label{tab:jingdu}
\resizebox{.9\textwidth}{!}{
\begin{tabular}{l|cccccc|cccccc} 
\toprule
\multicolumn{1}{c|}{} & \multicolumn{6}{c|}{10K Points}                                                                                                      & \multicolumn{6}{c}{5K Points}                                                                                                 \\ 
\cmidrule{2-13}
\multicolumn{1}{c|}{} & \multicolumn{2}{c}{NC~$\uparrow$} & \multicolumn{2}{c}{CD~$\downarrow$} & \multicolumn{2}{c|}{F1~$\uparrow$} & \multicolumn{2}{c}{NC~$\uparrow$} & \multicolumn{2}{c}{CD~$\downarrow$} & \multicolumn{2}{c}{F1~$\uparrow$}  \\
\multicolumn{1}{c|}{} & mean           & std.                    & mean          & std.                     & mean           & std.                   & mean           & std.                    & mean          & std.                     & mean           & std                    \\ 
\midrule
SPR~\footnotesize{\cite{kazhdan2013screened}}               & 94.16          & \textbf{4.48}                    & 7.62          & \textbf{3.05}                     & 74.54          & 26.65                  & 93.06          & 9.31                    & 9.36          & 6.55                     & 62.91          & 30.07                  \\ 
NSH~\footnotesize{\cite{wang2023neural}}             & 84.25   & 8.39    & 7.91     & 7.82    & 87.60     & 18.04                  
                                                     &83.85    & 11.28   & 9.78     & 17.48   & 82.61     & 18.99                  \\
NeurCAD~\footnotesize{\cite{dong2024neurcadrecon}}   & 89.92   & 7.33    & 6.70     & 7.22    & 89.44     & 17.50                  & 
                                                       84.15   & 10.28   & 8.79     & 15.57   & 83.42     & 17.91                  \\
NN-VIPSS~\footnotesize{\cite{ju2025variational}}    & 94.11     & 7.3         & 6.9         & 4.71    & 90.60     & 17.09                 & 90.35          & 9.28                    & 8.78         & 13.48                     & 85.61          & 18.32                  \\
BP~\footnotesize{\cite{bernardini1999ball}}                 & 92.26          & 9.24                    & 7.56         & 5.25                    & 82.95          & 22.23                  & 91.75          & 5.45                    & 8.76         & 9.71                    & 76.02          & 16.92                  \\
Greedy~\footnotesize{\cite{10.1007/s00371-003-0217-z}}           & 95.29          & 4.57                    & 7.43          & 5.15                     & 86.67          & 10.75        & 92.52          & 8.51                    & 7.97          & 10.37            & 82.32          & 17.74          \\
PTN~\footnotesize{\cite{sharp2020pointtrinet}}                   & \textbf{96.14}          & 6.12                   & 6.51         & 4.68                    & 89.47          & 14.06                  & \textbf{94.69}          & 17.97                   & 7.44         & 4.37                     & 87.14          & 17.57                  \\
RD~\footnotesize{\cite{wang2022restricted}}                 & 94.48          & 6.12                    & 6.71          & 6.94                     & 91.22          & 12.01                  & 92.55          & 7.11                    & 7.91          & 6.96                     & 89.86          & \textbf{11.52}                  \\

\midrule
\textbf{Ours}                  & 95.57 & 5.01           & \textbf{3.47}  & 4.41            & \textbf{93.63} & \textbf{10.05}                & 94.29 & \textbf{4.55}           & \textbf{4.12} & \textbf{3.46}                  & \textbf{90.34} & 12.17                  \\
\bottomrule
\end{tabular}
}
\end{table*}

\begin{figure*}[ht]
  \begin{center}
  \includegraphics[width=0.95\textwidth]{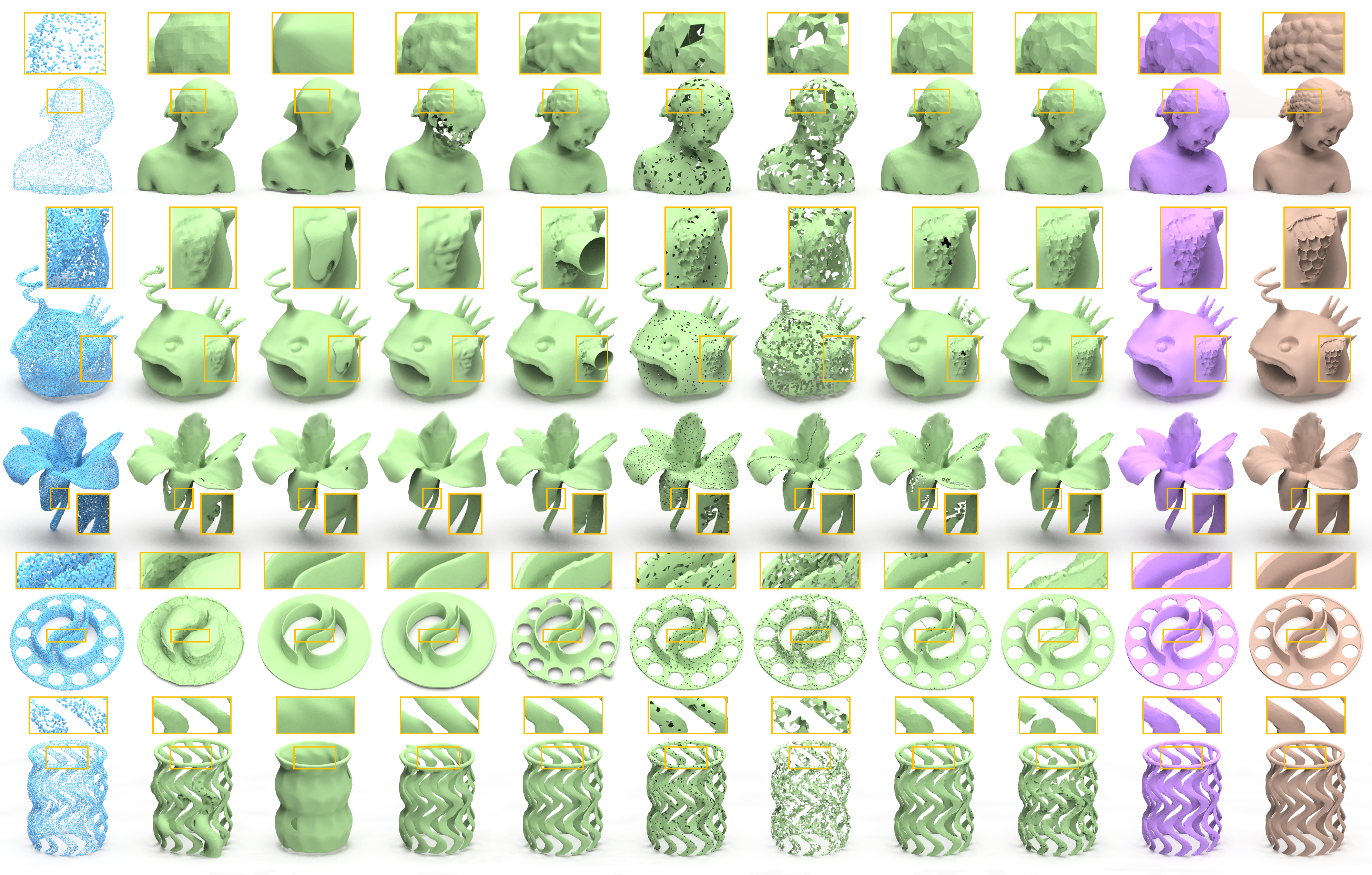}
  \makebox[0.082\textwidth][c]{\small (a) Input}
    \makebox[0.082\textwidth][c]{\small (b) SPR}
    \makebox[0.082\textwidth][c]{\small (c) NSH}
    \makebox[0.082\textwidth][c]{\small (d) NeurCAD}
    \makebox[0.082\textwidth][c]{\small (e) NN-VIPSS}
    \makebox[0.082\textwidth][c]{\small (f) PTN}
    \makebox[0.082\textwidth][c]{\small (g) BP}
    \makebox[0.082\textwidth][c]{\small (h) Greedy}
    \makebox[0.082\textwidth][c]{\small (i) RD}
    \makebox[0.082\textwidth][c]{\small (j) Ours}
    \makebox[0.082\textwidth][c]{\small (k) GT}
  \vspace{-5mm}
  \end{center}
   \caption{Comparing the reconstruction approaches. The input point clouds are sampled on thingi10k by Montecarlo sampling, the number of points is 5K - 20K.
   In terms of their ability to handle sharp features, thin plates, and geometric details. 
   Our method can recover the sharp features, thin-plate and geometric details.}
   \label{fig:gallery_thingi10k}
   \vspace{-2mm}
\end{figure*}

\subsection{Accuracy}
We employ multiple metrics to evaluate the accuracy of our reconstruction result: normal consistency, Chamfer distance, and the F-Score. Normal consistency, abbreviated as NC and expressed as a percentage, quantifies the alignment between the normals of the reconstructed and ground-truth surfaces. 
To compute the Chamfer distance, we normalize the values by a factor of $10^3$ and use it to gauge the closeness of fit between the two surfaces. 
The F-Score, also expressed as a percentage, serves as a balanced measure between precision and recall (or completeness). 
We set the default threshold for the F-Score to 0.005. 
Before evaluation, all meshes undergo uniform scaling to fit within the range of $[-0.5, 0.5]$.
For a more precise evaluation, we randomly selected 500 models from the Thingi10K and sample 10K and 5K points using the Monte-Carlo sampling, normal consistency, Chamfer distance, and the F-Score are demonstrated in Table~\ref{tab:jingdu}.
Additionally, we also generated 5K - 20K points by Montecarlo sampling.
In our algorithm no failure case was found - all the outputs are watertight manifolds.
In particular, the visual comparison in Fig.~\ref{fig:gallery_thingi10k} shows that our method can recover the sharp features, thin-plate and geometric details.

\begin{figure*}[ht]
  \begin{center}
  \includegraphics[width=0.95\textwidth]{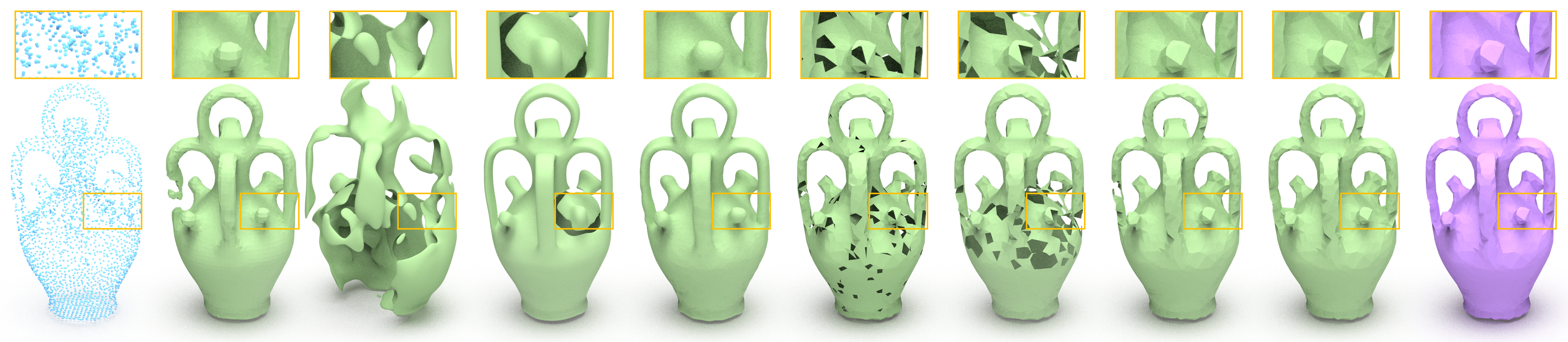}
  \makebox[0.092\textwidth][c]{\small   (a) Input}
    \makebox[0.092\textwidth][c]{\small (b) SPR}
    \makebox[0.092\textwidth][c]{\small (c) NSH}
    \makebox[0.092\textwidth][c]{\small (d) NeurCAD}
    \makebox[0.092\textwidth][c]{\small (e) NN-VIPSS}
    \makebox[0.092\textwidth][c]{\small (f) PTN}
    \makebox[0.092\textwidth][c]{\small (g) BP}
    \makebox[0.092\textwidth][c]{\small (h) Greedy}
    \makebox[0.092\textwidth][c]{\small (i) RD}
    \makebox[0.092\textwidth][c]{\small (g) Ours}
  \vspace{-4mm}
  \end{center}
   \caption{Irregular and sparse distribution of Point cloud. Input point cloud is synthesized sampling by~\cite{wang2022restricted}. 
   The sampling by employing Poisson disk, Monte Carlo sampling, and regular recursive sampling in the upper, middle, and bottom parts, resulting in significant variations in point density from top to bottom. 
   The limited quality of point distribution poses challenges in estimating normal vectors accurately, leading SPR to produce a surface with uneven bumps and broken handles. NSH and NeurCAD fail to generate correct topology. PTN struggles with sparse points, often resulting in a non-manifold and non-watertight mesh. BP demands a global ball-radius parameter, but determining an appropriate value for point clouds with varying densities remains challenging. Greedy fails to predict accurate connections between points in thin regions, causing broken handles. It is worth noting that NN-VIPSS excels at handling sparse and irregular point distributions, achieving an ideal result.
   While the majority of existing methods fail with low-quality input, our algorithm consistently exhibits success.
   }
   \label{fig:nonuniform_base}
   \vspace{-2mm}
\end{figure*}

\subsection{Irregular and Sparse Distribution of Point Cloud}
A point cloud is said to have a regular distribution if there is an almost even gap between points.
It's known blue-noise sampling can generate a set of points with a regular distribution.
The problem of surface reconstruction becomes difficult if the input points have an irregular and sparse distribution.

In Fig.~\ref{fig:nonuniform_base}(a), we using synthesized sampling by~\cite{wang2022restricted}. 
The sampling by employing Poisson disk, Monte Carlo sampling, and regular recursive sampling in the upper, middle, and bottom parts, resulting in significant variations in point density from top to bottom. 
Given the presence of thin tubes and plates in the Vase model, we subject various reconstruction approaches to rigorous testing to assess their ability to handle sparse, irregular, and non-uniform point cloud distributions. 
Our findings reveal that our algorithm effectively recovers the underlying geometry in this challenging scenario, delivering a watertight manifold mesh. 
In contrast, the outcomes of other algorithms exhibit notable flaws. 
The limited quality of point distribution poses challenges in estimating normal vectors accurately, leading SPR to produce a surface with uneven bumps and broken handles. 
NSH and NeurCAD fail to generate correct topology.
PTN struggles with sparse points, often resulting in a non-manifold and non-watertight mesh. 
BP demands a global ball-radius parameter, but determining an appropriate value for point clouds with varying densities remains challenging. 
Greedy fails to predict accurate connections between points in thin regions, causing broken handles.
It is worth noting that NN-VIPSS excels at handling sparse and irregular point distributions, achieving an ideal result.

\begin{figure*}[ht]
  \begin{center}
  \includegraphics[width=0.99\textwidth]{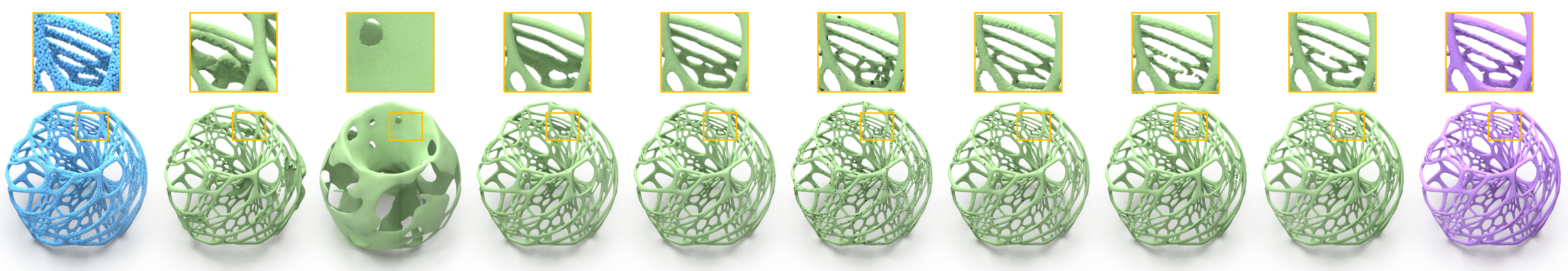}
  \makebox[0.092\textwidth][c]{\small   (a) Input}
    \makebox[0.092\textwidth][c]{\small (b) SPR}
    \makebox[0.092\textwidth][c]{\small (c) NSH}
    \makebox[0.092\textwidth][c]{\small (d) NeurCAD}
    \makebox[0.092\textwidth][c]{\small (e) NN-VIPSS}
    \makebox[0.092\textwidth][c]{\small (f) PTN}
    \makebox[0.092\textwidth][c]{\small (g) BP}
    \makebox[0.092\textwidth][c]{\small (h) Greedy}
    \makebox[0.092\textwidth][c]{\small (i) RD}
    \makebox[0.092\textwidth][c]{\small (g) Ours}
  \vspace{-4mm}
  \end{center}
   \caption{Comparing the ability to reconstruct the complicated high-genus tubular network. The input point cloud from ~\cite{wang2022restricted}, utilize the Poisson disk method to sample 50K on Thingi10K. See the highlighted windows, only our method reconstructed the correct topological in the thin tube region. }
   \label{fig:highGenus}
   \vspace{-2mm}
\end{figure*}

\subsection{High-genus, Thin Plates, and Geometric Details}
\paragraph{High Genus.} 
In Fig.~\ref{fig:highGenus}, we evaluate the performance of various algorithms using a complex, high-genus model. 
The input point cloud from ~\cite{wang2022restricted}, utilize the Poisson disk method to sample 50K on Thingi10K dataset. 
The implicit reconstruction methods tend to create undesired bumps, whereas the prevalent interpolation-based techniques often result in misconnections between gaps or broken branches in extremely thin tube regions. 
In contrast, our algorithm effectively interpolates the point cloud, ensuring a watertight manifold mesh that closely approximates the actual geometry, as shown in zoom of Fig.~\ref{fig:highGenus}.

\paragraph{Thin Plates, Geometric Details.}

\begin{figure*}[ht]
  \begin{center}
  \includegraphics[width=0.95\textwidth]{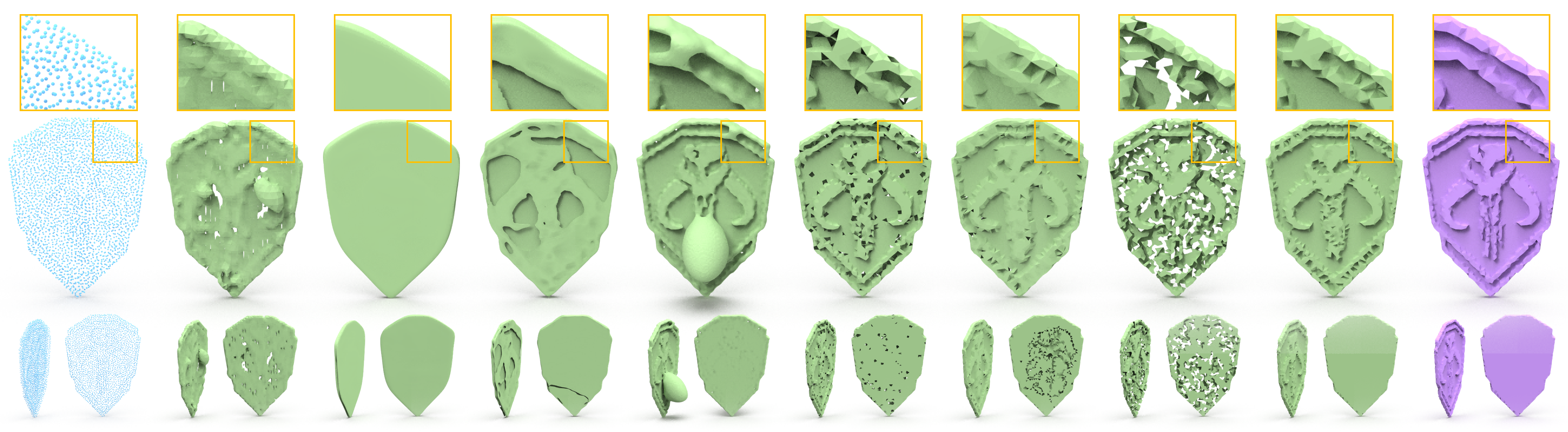}
  \makebox[0.092\textwidth][c]{\small   (a) Input}
    \makebox[0.092\textwidth][c]{\small (b) SPR}
    \makebox[0.092\textwidth][c]{\small (c) NSH}
    \makebox[0.092\textwidth][c]{\small (d) NeurCAD}
    \makebox[0.092\textwidth][c]{\small (e) NN-VIPSS}
    \makebox[0.092\textwidth][c]{\small (f) PTN}
    \makebox[0.092\textwidth][c]{\small (g) BP}
    \makebox[0.092\textwidth][c]{\small (h) Greedy}
    \makebox[0.092\textwidth][c]{\small (i) RD}
    \makebox[0.092\textwidth][c]{\small (g) Ours}
  \vspace{-4mm}
  \end{center}
   \caption{Comparing the ability to reconstruct the thin plates. The input point cloud from ~\cite{wang2022restricted}, utilize the Monte Carlo method to sample 3K. The thin-plate model has many geometric details on the front while remaining flat on the back. We give two separate views for each result. Our algorithm effectively distinguishes between the front and backside while maintaining intricate geometric details.}
   \label{fig:Thickness}
   \vspace{-2mm}
\end{figure*}

We conduct tests on thin models to assess the performance of various algorithms in handling thin plates and preserving geometric details. The Shield model by~\cite{wang2022restricted} (comprising 3K points sampled using the Monte Carlo method), reveals that our algorithm effectively distinguishes between the front and backside while maintaining intricate geometric details, see Fig.~\ref{fig:Thickness} zoom.
This is in contrast to SPR, NN-VIPSS, PTN, and Greedy, which tend to create numerous small and unintended topological holes. Additionally, it is noteworthy that NSH and NeurCAD tends to smooth out positive details, highlighting a key difference in its approach compared to our method.

\subsection{Real Scans}
For real scans, there are various defects such as noise, outliers, and data missing.
We test the performance of our method and six other methods on the real scanning datasets~\cite{huang2022surface}. 
At the same time, we also test the performance of our method on the Statue Model dateset~\cite{EPFL}, large-sized datasets~\cite{aanaes2016large}, and real scan by SHINING 3D Einscan SE scanner.

\begin{figure*}[ht]
  \begin{center}
  \includegraphics[width=0.95\textwidth]{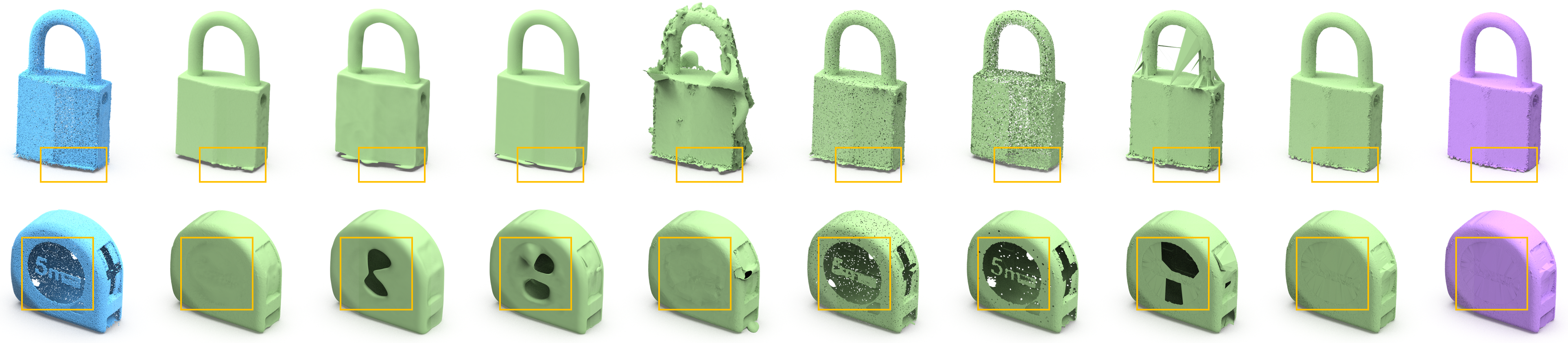}
  \makebox[0.092\textwidth][c]{\small   (a) Input}
    \makebox[0.092\textwidth][c]{\small (b) SPR}
    \makebox[0.092\textwidth][c]{\small (c) NSH}
    \makebox[0.092\textwidth][c]{\small (d) NeurCAD}
    \makebox[0.092\textwidth][c]{\small (e) NN-VIPSS}
    \makebox[0.092\textwidth][c]{\small (f) PTN}
    \makebox[0.092\textwidth][c]{\small (g) BP}
    \makebox[0.092\textwidth][c]{\small (h) Greedy}
    \makebox[0.092\textwidth][c]{\small (i) RD}
    \makebox[0.092\textwidth][c]{\small (g) Ours}
  \vspace{-4mm}
  \end{center}
   \caption{Real scan point cloud with noise (top) and data missing (bottom). }
   \label{fig:realscan_noise_datamissing}
   \vspace{-2mm}
\end{figure*}

\paragraph{Noise and Data Missing}
The tests performed on the Lock and Tape model, as cited in \cite{huang2022surface}, clearly indicate that a significant number of existing algorithms falter in the face of noise and data loss. RD and our own algorithm, on the other hand, demonstrate remarkable resilience to noise and an ability to fill in gaps, resulting in a robust, watertight manifold mesh, without introducing unnecessary topological complexities. 
Fig.~\ref{fig:realscan_noise_datamissing} offers illustrative examples of challenging scenarios involving noise and missing data. Nevertheless, under extreme data sparsity, where the point cloud barely resembles a coherent surface, our algorithm may struggle to produce meaningful results.

When dealing with severe noise, the task of accurately estimating normal directions becomes particularly challenging. 
This, in turn, severely undermines the quality of the scalar field underlying implicit surface reconstruction techniques. Our algorithm, uniquely, does not rely on predefined normal vectors. Instead, it progressively deduces the true normals as it refines the guiding surface. 
Our method can easily resolve the topological issue,
thus naturally eliminating the adverse influence of distant noisy points, as shown in Fig.~\ref{fig:realscan_noise_datamissing}.

SPR and NN-VIPSS excels at seamlessly filling in missing data regions with smoothly interpolated surfaces. 
NSH and NeurCAD generate an incorrect double-layer surface. 
PTN, despite incorporating soft penalties to enhance the production of manifold and watertight surfaces, still falls short when confronted with significant data loss. 
BP requires precise parameter tuning to accommodate large gaps between points, and if the data loss is too extensive, these gaps may remain unfilled. 
Greedy approaches tend to fill empty spaces with oversized triangles, sacrificing both manifold guarantees and outlier management. 
In contrast, both RD and our algorithm are uniquely equipped to tackle a wide range of defects simultaneously.

We also tested our algorithm on the Statue Model dateset~\cite{EPFL} shown in Fig.~\ref{fig:teaser} brown model, which contains noise and data loss. 
It can be seen that our algorithm can deal with these artifacts.

\paragraph{Large-size Point Clouds}
There also exists a dataset~\cite{aanaes2016large} that encompasses extensive point clouds obtained from scanning 40 genuine objects. It is crucial to evaluate our algorithm's ability to deal with large-size point clouds. For rigorous testing, we intentionally select a single-view stereo scan for each model, with sizes ranging from 1000K to 4000K points. 
Our algorithm necessitates approximately 3 hours to complete the reconstruction process. 
A subset of these results is showcased in Fig.~\ref{fig:teaser} brown model.

\begin{figure}[!t]
  \begin{center}
  \includegraphics[width=0.95\columnwidth]{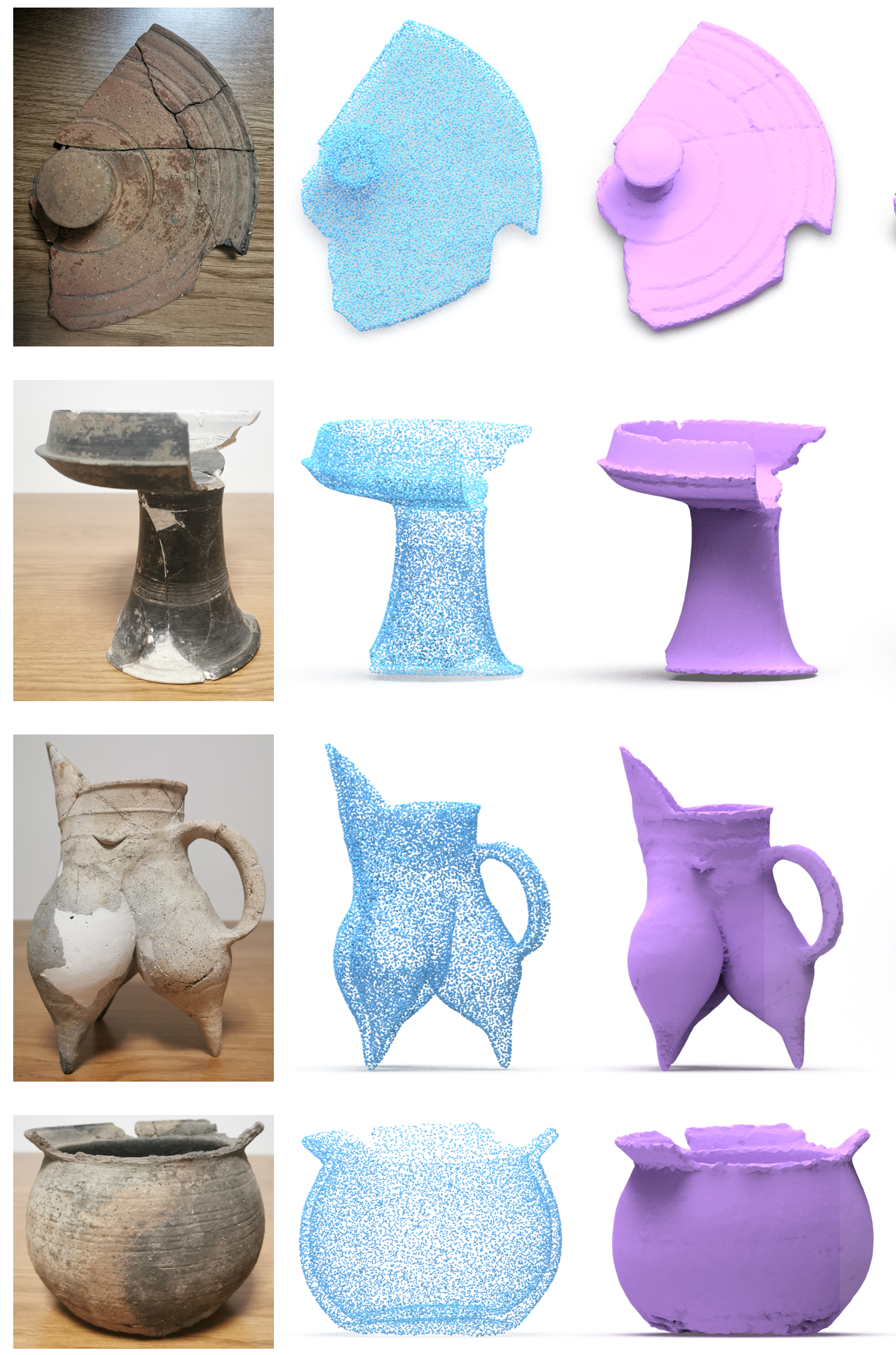}
  \makebox[0.155\textwidth][c]{\small (a) Real Objects }
     \makebox[0.155\textwidth][c]{\small (b) Point clouds }
      \makebox[0.155\textwidth][c]{\small (c) Results}
  \vspace{-4mm}
  \end{center}
    \caption{The results of our approach on real scanned point clouds. Reconstruction results for point clouds scanned by the SHINING 3D Einscan SE scanner.}
   \label{fig:real_scan_bowuguan}
   \vspace{-2mm}
\end{figure}

\paragraph{Real scan by Laser Scanner}
To this end, we also evaluate our method’s performance on real scanned models. 
The point clouds generated by the scanner, specifically captured using a SHINING 3D Einscan SE scanner with an accuracy of 0.1mm, present challenges such as missing and nonuniform density. 
For our analysis, the scanned point clouds contain over 20K points.
As depicted in the upper part of Fig.~\ref{fig:real_scan_bowuguan}, our method demonstrates a remarkable ability to recover fine details
and accurately render concave regions of shapes, showcasing its effectiveness in dealing with real-world model reconstructions.
Notably, our method demonstrates robustness to data missing and nonuniform, consistently yielding faithful reconstructions of models.

\begin{table}[!t]
\centering
\caption{ Runtime performance statistics.}
\label{tab:time}
\resizebox{\linewidth}{!}{
\begin{tabular}{l|ccc|ccc} 
\toprule
\multicolumn{1}{c|}{} & \multicolumn{3}{c|}{RD~\cite{wang2022restricted}}                                                                                                      & \multicolumn{3}{c}{Ours}                                                                                                 \\ 
\cmidrule{2-7}
 & Filmsticking &Sculpting & Time(s) & Filmsticking++ & Poisson reconstruction & Time(s)  \\

\midrule
Girl (5K) Fig.~\ref{fig:gallery_thingi10k}               & 5                    & 3                         & 2.6                      & 4                      & 1                             & 1.9                      \\ 

Flower (20k) Fig.~\ref{fig:gallery_thingi10k}               & 20                         & 10                          & 17.94                         & 8                        & 1          & 3.69                                    \\

Disk (20K) Fig.~\ref{fig:gallery_thingi10k}                  & 10           & 7              & 10.75               & 7         & 1                 & 3.1                   \\
Dragon (150K) Fig.~\ref{fig:teaser}      & -           & -             & -               & 17         & 1                 & 698                   \\
\bottomrule
\end{tabular}
}
\end{table}

\subsection{Run-time Performance Statistics and Analysis}
Suppose that the input point cloud $\mathbf{P}$ contains $n$ points and finally produces a triangle mesh of about $2n$ triangle faces. 
For efficient query of the distance from a point to~$\mathbf{P}$, we pre-build a kd-tree of~$\mathbf{P}$ in $O(n\log n)$ time~\cite{arya1998ann} and pre-compute the Delaunay triangulation w.r.t.~$\mathbf{P}$ in~$O(n\log n)$ time in general condition~\cite{attali2003complexity}.
Throughout the algorithm, the guiding surface $\mathbf{G}$ has a complexity of not exceeding $2n$ faces. 
Our algorithm generally requires only a few iterations, typically 10 to 20 Filmsticking++ and one Poisson reconstruction. 
Considering that either computing RPD once on $\mathbf{G}$ or Poisson reconstruction once requires about $O(n)$ time,
the empirical time complexity can be estimated by~$O(n\log n) + kO(n)$,
where $k$ equals to the number of computing RPDs plus one Poisson reconstruction. 
Empirical evidence shows that $k$ is smaller than 20. 
The processing times for the models in Fig.~\ref{fig:gallery_thingi10k} and Fig.~\ref{fig:teaser} are listed in Table~\ref{tab:time}. It is important to note that if the number of points is excessively large, both the Filmsticking++ step and the Poisson reconstruction process will become significantly time-consuming.

\section{Conclusions and Limitations}
Inspired by~\cite{wang2022restricted}, we introduce a more rapid, flexible, and scalable filmsticking method, named~\textit{Filmsticking++}, to reconstruct the interpolation-based surface, assuming a poor-quality, unoriented point cloud as input, while the target surface may exhibit topologically and geometrically complicated shapes. Filmsticking++ improves the state-of-the-art (SOTA) method in several ways: (1) We propose replacing the restricted Voronoi diagram with the restricted power diagram, allowing more points to be attracted onto the guiding surface during each iteration; (2) We cast a set of randomly distributed points, typically around 1K, within the space of interest (e.g., the interior of the initial guiding surface). These points assist the guiding surface in rapid stretching into the deep internal cavities; and (3) we utilize an field-based method only one time, which is used to eliminate thin topological plates. Compared with the SOTA method, Filmsticking++ exhibits several advantages, including fewer iterations, improved robustness, and better scalability.

Our current algorithm cannot handle the reconstruction of open models due to an overestimation of the minimum thickness. Additionally, if the target surface includes a very thin neck, the inward offset may break into disconnected components, thereby resulting in an incorrect guiding surface. In the future, we shall explore the possibility of using a spatially varying thickness parameter to address these issues.

\bibliographystyle{ACM-Reference-Format}
\bibliography{main}


\end{document}